\documentclass[10pt]{article}
\pdfoutput=1
\usepackage{jheparxiv}
\usepackage[latin1]{inputenc}
\usepackage{latexsym,diagbox,stackrel}
\usepackage{relsize}
\usepackage{shuffle}
\usepackage{amsmath}
\usepackage{mathrsfs}
\usepackage{dsfont}
\usepackage{array}
\usepackage{nicefrac}
\usepackage{arydshln,leftidx,mathtools}
\usepackage{bbold}
\usepackage{graphicx}
\usepackage{subcaption}
\usepackage{pdflscape}
\usepackage{amssymb}
\usepackage{multirow}
\usepackage{enumerate}
\usepackage{blkarray}
\usepackage[dvipsnames,svgnames,table]{xcolor}
\usepackage{cleveref}
\usepackage{tikz}
\usepackage{graphicx}
\usepackage{slashed}
\usetikzlibrary{shapes.geometric, arrows}
\definecolor{light-gray}{gray}{0.95}
\definecolor{light-grayII}{gray}{0.85}
\usepackage{array}
\usepackage{pdfpages}
\usepackage{empheq}
\usepackage{comment}
\usepackage{graphbox}

\newcommand{\C}{\mathbb{C}}

\newcommand{\cI}{\mathcal{I}}

\newcommand{\SL}{\mathrm{SL}}

\newcommand{\tr}{\mathrm{tr}}

\newcommand{\vol}{\mathrm{vol}\,}

\renewcommand{\[}{\begin{equation}\begin{aligned}}
\renewcommand{\]}{\end{aligned}\end{equation}}

\newcommand*\widefbox[1]{\fbox{\hspace{2em}#1\hspace{2em}}}

\subheader{\hfill \begin{tabular}{r} \texttt{QMUL-PH-24-23} \end{tabular}}
 
\title{\Large Superstring amplitudes from BCJ numerators at one loop}
\author[a]{Yvonne Geyer,}
\author[b]{Jiachen Guo,}
\author[b]{Ricardo Monteiro}
\author[b]{\& Lecheng Ren}
        
\affiliation[a]{Department of Physics, Faculty of Science, Chulalongkorn University,\\
Thanon Phayathai, Pathumwan, Bangkok 10330, Thailand}

\affiliation[b]{Centre for Theoretical Physics, Department of Physics and Astronomy, \\
        Queen Mary University of London, E1 4NS, United Kingdom}
        
\emailAdd{yjgeyer@gmail.com}
\emailAdd{jiachen.guo@se22.qmul.ac.uk}
\emailAdd{ricardo.monteiro@qmul.ac.uk}
\emailAdd{lecheng.ren@qmul.ac.uk}

\abstract{
We find a direct map that determines moduli-space integrands for one-loop superstring amplitudes in terms of field-theory loop integrands in the BCJ form. The latter can be computed using efficient unitarity methods, so our map provides an alternative to worldsheet CFT techniques. This construction is a one-loop higher-point analogue of a recent conjecture for the three-loop four-point superstring amplitude. Based on the one-loop chiral-splitting representation, we show how all the coefficients of an ansatz for the superstring can be identified with field-theory BCJ numerators, up to at least 7-point amplitudes. Moreover, we obtain partial results for all higher-point amplitudes. The monodromy constraints associated to chiral splitting play a crucial role in determining coefficients of the ansatz that, naively, are not fixed by the field-theory limit. Taking a field-theory perspective, our ansatz for the superstring implies by construction the existence of one-loop BCJ numerators at any multiplicity.
}

\begin{document}

\maketitle


\section{Introduction}

Is perturbative string theory, at least for the scattering of massless states, in some particular sense just an $\alpha'$-dressing of perturbative field theory? To provide a positive answer to this question, one would have to find a procedure whereby all the information needed to construct a string theory amplitude could be extracted from its field-theory limit. It is far from obvious that such a procedure can provide all the information, as we will discuss, because something may be lost in the limit. On the other hand, if the answer is positive, this represents an interesting alternative to the worldsheet conformal field theory, at least when performing practical calculations. As powerful as the 2D CFT can be, computing loop-level string amplitudes remains a formidable problem, not just at the level of the moduli-space integral, but even at the level of the moduli-space integrand, i.e.~the CFT correlator; see ref.~\cite{Berkovits:2022ivl} for a recent review.

Our reasoning has already led to an explicit conjecture for the 3-loop 4-point superstring amplitude \cite{Geyer:2021oox}. Instead of employing traditional worldsheet techniques, like the RNS \cite{DHoker:1988pdl,DHoker:2002hof,Witten:2012bh} or the pure spinor \cite{Berkovits:2000fe,Berkovits:2002zk} formalisms, the goal was to bootstrap the moduli-space integrand purely from the knowledge of the field-theory limit. While the field-theory amplitude is not defined in 10D, the loop integrand is, and there has been tremendous progress in determining higher-loop field-theory integrands using unitarity techniques and also the double copy, with a wide range of applications \cite{Bern:2011qt,Bern:2019prr,Adamo:2022dcm}. Our main goal in this paper is to show that the idea exploited in \cite{Geyer:2021oox} can be applied more broadly to produce results for string correlators that are structurally simple. Here, we apply the idea to one-loop superstring amplitudes, but at higher multiplicity.

Our approach relies, of course, on the many lessons provided by previous work on the RNS and pure-spinor formalisms. In order to construct an ansatz for the superstring correlator (whose coefficients we will bootstrap using the field-theory limit), it is essential to understand the space of worldsheet objects that may enter the correlator. It turns out that, for our purposes, it is convenient to employ the chiral-splitting representation of the superstring amplitudes \cite{DHoker:1989cxq,DHoker:1988ta}, such that the moduli-space integrand is expressed as a `double copy' at fixed loop momentum. Our description of the chirally-split one-loop correlators and their modularity/monodromy properties builds strongly on the works \cite{Broedel:2014vla,Mafra:2017ioj}; see also
\cite{Mafra:2012kh,Mafra:2018nla,Mafra:2018pll,Mafra:2018qqe,Rodriguez:2023qir,Zhang:2024yfp}. We expect that some form of the procedure applies at any loop order. In fact, although the main focus of \cite{Geyer:2021oox} was on the 3-loop 4-point amplitude, the 2-loop 4-point amplitude was also dealt with, easily reproducing the known result \cite{DHoker:2005vch,Berkovits:2005df,Berkovits:2005ng}. We should note that the simplest part of the 3-loop 4-point correlator constructed in \cite{Geyer:2021oox} had been previously obtained using the pure-spinor formalism \cite{Gomez:2013sla}.

Chiral splitting is the string theory manifestation of the double copy, going back to the tree-level KLT relations \cite{Kawai:1985xq}. In the field-theory limit, the double copy has been the basis for dramatic progress in our understanding of scattering amplitudes for gauge theory and gravity; see \cite{Bern:2019prr,Adamo:2022dcm} for recent reviews. The BCJ double copy \cite{Bern:2008qj,Bern:2010ue,Bern:2017yxu} has been particularly useful at loop level. Its relation to string theory has been widely studied, e.g.~\cite{BjerrumBohr:2009rd,Stieberger:2009hq,Mafra:2011kj,Ochirov:2013xba,Mafra:2014gja,He:2015wgf,Mafra:2015vca,Tourkine:2016bak,Hohenegger:2017kqy,Ochirov:2017jby,Tourkine:2019ukp,Mizera:2019gea,Casali:2019ihm,Casali:2020knc,Borsten:2021hua,Bridges:2021ebs}, and it seems to be particularly clear in the pure-spinor formalism \cite{Mafra:2011kj,Mafra:2022wml,Ben-Shahar:2021doh}. In this one-loop paper, we will not rely on any particular formalism, but will instead show that the mathematical structure of the superstring correlator not only encodes, but is encoded by the BCJ structure of the field-theory limit, at least up to 7 points (and partly at any multiplicity). We note that chiral splitting has also been recently used to describe the one-loop KLT relations \cite{Stieberger:2023nol,Stieberger:2022lss,Bhardwaj:2023vvm,Mazloumi:2024wys}, again showing its relevance to the double copy.

The main idea behind our paper and also its predecessor \cite{Geyer:2021oox} originated in a worldsheet formulation of field theory, and was then imported into string theory. By worldsheet formulation of field theory, we mean the twistor \cite{Witten:2003nn} and ambitwistor \cite{Mason:2013sva} strings, and the associated formulas for amplitudes based on the scattering equations, at tree level \cite{Roiban:2004yf,Cachazo:2013hca,Cachazo:2013iea,Geyer:2014fka} and at loop level \cite{Adamo:2013tsa,Casali:2014hfa,Adamo:2015hoa,Geyer:2015bja,He:2015yua,Geyer:2015jch,Cachazo:2015aol,Geyer:2016wjx,Geyer:2017ela,Geyer:2018xwu,Geyer:2019hnn}; see also \cite{Berkovits:2013xba,Baadsgaard:2015voa,Feng:2016nrf,Cardona:2016bpi,Cardona:2016gon,Chen:2016fgi,Gomez:2016cqb,Gomez:2017lhy,He:2017spx,Gomez:2017cpe,Roehrig:2017gbt,Ahmadiniaz:2018nvr,Edison:2020uzf,Abhishek:2020sdr,Kalyanapuram:2021xow,Kalyanapuram:2021vjt,Feng:2022wee,Dong:2023stt,Xie:2024pro}. Ref.~\cite{Geyer:2022cey} presents a review of these developments. In fact, the chiral integrands that we will obtain, which enter into the chiral-splitting representation of the superstring amplitude, are also valid for the ambitwistor string.

We want to emphasise a particular approach that we will use in dealing with the field-theory limit of the one-loop superstring, already employed in \cite{Geyer:2021oox} at two and three loops. This limit has mostly been studied using `pinching rules', whereby the field-theory Feynman diagrams are obtained from the degeneration of the Riemann surface, namely the torus at one loop \cite{Green:1982sw,Bern:1991aq,Strassler:1992zr,Schubert:2001he,Tourkine:2013rda}. The lesson of the ambitwistor/CHY story, however, is that it may be better to not look individually at each diagram, but instead to consider the complete amplitude or integrand. This is the approach that we will follow.

While ambitwistor strings played an important role in unveiling the tools we employ, our paper will focus solely on the conventional superstring. What is surprising to us in this whole story is that -- to this date, and going back to the beginning of this Introduction -- the superstring loop amplitude appears to be `merely' an $\alpha'$-regularisation of field theory, without changing essential integrand-level features. Our results clearly illustrate this observation, but also refine the conditions for its potential failure at higher points or loops.

This paper is organised as follows. In section~\ref{sec:structure}, we review chiral splitting, describe how our ansatze for superstring correlators are constructed, and discuss the relation to the BCJ double copy. After a comment on notation in section~\ref{sec:symmetrise}, we present our map relating the chiral integrands and the BCJ numerators up to 7 points, in sections~\ref{sec:4pt} to \ref{sec:7pt}. We discuss the part of the $n$-point structure that can already be unveiled in section~\ref{sec:npt}. In section~\ref{sec:checks}, we illustrate the results with explicit formulas for the BCJ numerators, and perform checks of gauge invariance. Finally, we discuss the conclusions and future directions in section~\ref{sec:conclusion}. We also include some details of the 6-point discussion in Appendix~\ref{app:6ptBCJmonodromies}.

{\bf Note Added:} We call the readers' attention to the closely related work \cite{Balli:2024wje}, which develops systematic pinching rules to obtain the field-theory limit of one-loop superstring amplitudes in the chiral-splitting representation. In our paper, the analogue of using these rules is the straightforward -- but, at this stage, {\it assumed} -- relation between the numerators in \eqref{eq:assumpq0} and those in \eqref{eq:Aft}. In addition, our results are consistent with those of \cite{Balli:2024wje}, in that the form of the 6-point superstring amplitude found there fits our 6-point ansatz \eqref{eq:I6initial}. In particular, terms with $\partial_{z_i}g^{(1)}_{ij}$, absent from our ansatz but appearing in previous literature on superstring correlators \cite{Mafra:2018pll}, were verified in \cite{Balli:2024wje} to not be essential, due to integration by parts. We are grateful to the authors of \cite{Balli:2024wje} for notifying us of their results, and for coordinating the submission.

\section{Structure of one-loop superstring amplitude}
\label{sec:structure}

Our starting point is the chiral-splitting formalism for one-loop superstring amplitudes \cite{DHoker:1988ta,DHoker:1989cxq}. In this formalism, one introduces an additional integration when writing the amplitude as a moduli-space integral, which will correspond in the field-theory limit to the usual loop integration. This additional integration achieves the holomorphic/antiholomorphic factorisation of the moduli space integrand. 
For the closed string, we have
\begin{equation}
\label{eq:Ac}
 \mathcal{A}_n^{(1),\text{closed}} = \alpha'^n \int_{\mathbb R^D} d^D\ell \int_{\mathcal{F}} d^2\tau \int_{T^{n-1}}\!d^2z_2\cdots  d^2z_n\;\;\mathcal{I}_n(\ell)\;\tilde{\mathcal{I}}_n(\ell)\;\big|\text{KN}_n(\ell)\big|^2 \,.
\end{equation}
Here, $\tau$ is the modular parameter, to be integrated over the fundamental domain of the modular group SL${}_2(\mathbb Z)$. Each of the integrations over the marked points $z_{i>1}$ cover the torus parallelogram with corners at $\{0,1,\tau+1,\tau\}$, with $z_1$ fixed using translation invariance as usual. The integration over $\ell$ corresponds in the field-theory limit to the loop integration. The integrand in \eqref{eq:Ac} is factorised into chiral and anti-chiral copies, associated to factorised external states (e.g.~$\varepsilon_i^{\mu\nu}=\epsilon_i^\mu\tilde\epsilon_i^\nu$). The quantity $\mathcal{I}_n(\ell)$ is our main object of study, while the chiral Koba-Nielsen factor is given by
\begin{equation}
\label{eq:KN}
 \text{KN}_n(\ell) = \exp\frac{\alpha'}{2}\Bigg(\sum_{1\leq i<j\leq n} p_i\cdot p_j\ln \theta_1(z_{ij},\tau) +2i\pi\,\ell\cdot\sum_{j=1}^n z_j\, p_j \,+i\pi\tau\,\ell^2\Bigg)\,.
\end{equation}
We define the odd Jacobi theta function as
\[
\theta_1(z,\tau) := 2q^{1/8} \sin(\pi z) \prod^\infty_{n=1} (1-q^n) (1-q^n e^{2\pi i z})(1-q^n e^{-2\pi iz})\,,
 \quad \text{with} \quad q:=e^{2\pi i\tau}\,.
\]
Similarly, for the open string we have
\begin{equation}
\label{eq:Ao}
 \mathcal{A}_n^{(1),\text{open}} = \alpha'^n \sum_\text{top} C_\text{top} \int_{\mathbb R^D} d^D\ell\int_{\mathcal{D}_\text{top}} d\tau \int_{O^{n-1}_\text{top}}\!d^2z_2\cdots  d^2z_n\;\;\mathcal{I}_n(\ell)\;\big|\text{KN}^{(\alpha'\mapsto4\,\alpha')}_n(\ell)\big| \,,
\end{equation}
where we sum over the cylinder and Moebius-strip topologies, with associated Chan-Paton colour factors $C_\text{top}$, and we integrate over the usual open-string domains for $\tau$ and the marked points $z_{i>1}$. We also noted that, to match standard open-string conventions, we need the substitution $\alpha'\mapsto4\,\alpha'$ in the Koba-Nielsen factor \eqref{eq:KN}. Comparing \eqref{eq:Ao} to \eqref{eq:Ac}, we see that there is an obvious `double copy'. It is no surprise that the chirally-split representation of the one-loop amplitudes is particularly well-suited to connect with advances in field theory related to the BCJ double copy \cite{Bern:2008qj,Bern:2010ue,Bern:2019prr}. We note that the $\ell$-integration is of Gaussian type, because of the form of the Koba-Nielsen factor \eqref{eq:KN} and of the fact that $\mathcal{I}_n(\ell)$ depends on $\ell$ polynomially. To avoid repetition, we refer mostly to the closed superstring in this paper, with the understanding that the same object $\cI_n$ can be imported into the open superstring.\footnote{In fact, the same object can also be imported into the heterotic string as the `supersymmetric copy'.}

The first step in the field-theory limit is to consider where $\alpha'$ occurs in the amplitude formula. We see an explicit $\alpha'^n$ normalisation factor in both \eqref{eq:Ac} and \eqref{eq:Ao}, and we note also the $\alpha'$ dependence of the Koba-Nielsen factor \eqref{eq:KN}. The question then is whether $\cI_n$ depends on $\alpha'$ for the superstring, given we already set the appropriate $\alpha'$ normalisation of the amplitude. If the superstring amplitude is an $\alpha'$-dressing of field theory, then there should be a valid choice of $\cI_n$ where there is no dependence on $\alpha'$. Note also that there is an ambiguity in the moduli-space integrand, given by total derivatives in $\cI_n \text{KN}_n$. These two observations are closely related. Indeed, in recent work parallel to ours, ref.~\cite{Balli:2024wje} was able to achieve a form of the 6-point chiral integrand that fits the ansatz we will present, precisely after taking into account integration by parts that gets rid of the $\alpha'$ dependence in a pure-spinor-derived formula for $\cI_6$.\footnote{Ref.~\cite{Balli:2024wje} corrects an integration-by-parts step in \cite{Mafra:2018pll} by appropriately requiring single-valuedness in the argument of total derivatives.} 
In what follows, we will assume that $\cI_n$ is independent of $\alpha'$ for the superstring, which is consistent with all present results.\footnote{A word on this assumption. The results mentioned include discussions at arbitrary genus on the relation to the ambitwistor string, where the limit $\alpha'\to\infty$ is taken to obtain $\cI_n^\text{ambitwistor}$ \cite{Mason:2013sva,Berkovits:2013xba,Huang:2016bdd,Casali:2016atr,Azevedo:2017yjy,Kalyanapuram:2021xow}, when one would expect the limit $\alpha'\to0$ because the ambitwistor string should represent field theory. Our view is that there should be a choice (or class of choices) where $\cI_n$ is independent of $\alpha'$, as we mentioned, and can then be used in both the superstring and the ambitwistor string. This appears to us to be the natural way to square the results at $\alpha'\to\infty$ and $\alpha'\to0$.}

\subsection{Ansatze for the chiral integrands}

Our goal in this paper is to construct the chiral integrand $\mathcal{I}_n$ in such a manner that the amplitude formula makes sense with respect to double periodicity and modular transformations. Following \cite{Broedel:2014vla,Mafra:2017ioj}, we will employ the functions $g^{(w)}$ generated from the Kronecker-Eisenstein series:
\begin{equation}
 F(z,\alpha, \tau):=  \frac{\theta_1'(0,\tau)\,\theta_1(z+\alpha,\tau)}{\theta_1(\alpha,\tau)\, \theta_1(z,\tau)} = \sum_{w=0}^\infty \alpha^{w-1}g^{(w)}(z,\tau)\,.
\end{equation}
We have $g^{(0)}(z,\tau) = 1$, and then
\begin{equation}
  g^{(1)}(z,\tau) = \partial_z\ln\theta_1(z,\tau)\,,\qquad 2g^{(2)}(z,\tau) = \big(\partial_z\ln\theta_1(z,\tau)\big)^2 +\partial_z^2 \ln\theta_1(z,\tau)-\frac{\theta_1'''(0,\tau)}{3\theta_1(0\tau)}\,,
\end{equation}
and so on. Notice that $g^{(1)}(z,\tau)$ has a simple pole at $z=0$, whereas $g^{(w)}(0,\tau)$ is regular for $w>1$. Notice also that $F(z,\alpha,\tau)$ is quasi-doubly-periodic, with
\begin{equation}
 F(z+1,\alpha,\tau) = F(z,\alpha,\tau)\,,\qquad F(z+\tau,\alpha,\tau)= e^{-2i\pi\alpha}F(z,\alpha,\tau)\,.
\end{equation}
Hence, the functions $g^{(w)}$ transform non-trivially under monodromies, as we will describe later. Moreover, the $g^{(w)}$ satisfy so-called Fay relations, obtained by expanding the following identity in $\alpha_1$ and $\alpha_2$:
\[
F(z_1,\alpha_1)F(z_2,\alpha_2) = F(z_1,\alpha_1+\alpha_2)F(z_2-z_1,\alpha_2)+F(z_2,\alpha_1+\alpha_2)F(z_1-z_2,\alpha_1)\,.
\]
For instance, at the lowest non-trivial order, we have the Fay relation\footnote{The notation, to be used also later, means that the expression includes cyclic permutations of the indices $\{1,2,3\}$.}
\begin{equation}
  g^{(1)}_{12} g^{(1)}_{23} + g^{(2)}_{13} + \text{cyc}(1, 2, 3) = 0\,.
\end{equation}
For brevity, here and in the following, we define
\[
 g^{(w)}_{ij} = g^{(w)}(z_i-z_j,\tau)\,.
\]
Notice also the parity property:
 \[
 g^{(w)}_{ij}=(-1)^w g^{(w)}_{ji}\,.
 \]
 
We keep track of the modular properties of the chiral integrand by associating a related weight to its building blocks, following \cite{Broedel:2014vla}. In particular, we require that the $n$-point chiral integrand $\mathcal{I}_n$ has weight $n-4$. We consider the following building blocks, consistent with meromorphicity.
\[
\label{eq:tablew}
\begin{tabular}{|c|c|c|c|}
\hline
object & weight \\
\hline
$2\pi i\,\ell_\mu$ & 1 \\
\hline
$ g^{(w)}_{ij}$ & $w$ \\
\hline
$G_{2K}$ & $2K$ \\
\hline
\end{tabular}
\]
Although it will not make an appearance in our $(n\leq 7)$-point examples, we included for completeness the holomorphic Eisenstein series. Setting $K\geq2$, \,$G_{2K}(\tau) = - g^{(2K)}(0,\tau)$\, are modular forms with weight $2K$; each can be written as a polynomial in $G_4$ and $G_6$ with appropriate weight. These objects may first appear in a superstring chiral integrand at 8 points, corresponding to $G_4$. We will make further comments on the appearance of these modular forms at the end of section~\ref{sec:npt}.

In fact, the objects $ g^{(w)}_{ij}$ do not have a well-defined modular weight, and are not doubly-periodic. Instead, those properties hold for their non-holomorphic cousins \cite{brown2013multipleellipticpolylogarithms},
\[
f^{(w)}(z,\tau) = \sum_{m=0}^w \frac1{m!}\left(2\pi i\;\frac{\text{Im}z}{\text{Im}\tau}\right)^m g^{(w-m)}(z,\tau)\,.
\]
It turns out (see e.g.~\cite{Mafra:2017ioj,Mafra:2018pll}) that
when the integration over $\ell$ is performed in \eqref{eq:Ac} and \eqref{eq:Ao}, the objects $ g^{(w)}_{ij}$ are effectively replaced by $ f^{(w)}_{ij}$, and the resulting moduli-space integrand becomes manifestly doubly-periodic. The required cancellations for this to occur rely on the property that the integrands in \eqref{eq:Ac} and \eqref{eq:Ao} are invariant under monodromy as one takes a marked point $z_i$ around a cycle, up to a shift in the loop momentum. The non-trivial constraint is the invariance under the operation
\[
\label{eq:monodromy}
i\text{-particle monodromy:} \qquad \ell^\mu \mapsto \ell^\mu - p_i^\mu\,, \qquad 
z_i\mapsto z_i+\tau\,.
\]
It is straightforward to show that the chiral Koba-Nielsen factor \eqref{eq:KN} is invariant. To impose the same property on the chiral integrand $\mathcal{I}_n$, we will use the identity
\[
g^{(w)}(z+\tau,\tau) = \sum_{m=0}^w \frac{(-2\pi i)^m}{m!}\, g^{(w-m)}(z,\tau)\,.
\]

Let us illustrate our choice of ansatze, using examples that we will later revisit in detail. The ansatz for the 5-point chiral integrand, which has weight 1, is
\[
\label{eq:5pteg}
\cI_5= 2\pi i \,C_5^\mu \,\ell_\mu + \sum_{\substack{i,j\\ i<j}}\, C_{5,ij}\, g^{(1)}_{ij}\,,
\]
where the $C$'s are functions only of the external particle data ($p_i^\mu$ and e.g.~$\epsilon_i^\mu$). At 6 points, the weight-2 ansatz for $\cI_6$ has contributions of the form $(2\pi i)^2\ell_\mu\ell_\nu$, $2\pi i\,\ell_\mu\, g^{(1)}$, $g^{(1)}\,g^{(1)}$ and $g^{(2)}$. And so on at higher points. There will be further restrictions on our ansatze related to closed cycles of $z_{ij}$, starting e.g.~with the absence of $g^{(1)}_{ij}g^{(1)}_{ji}$ at 6 points, which exhibits a double pole at $z_i=z_j$.

The complete chiral integrand $\cI_n \,\text{KN}_n$ is only defined up to total derivatives of single-valued functions, so $\cI_n$ is not unique. We will discuss in examples how gauge invariance is achieved in this way. In the field-theory limit to be discussed in the following, this will also be related to the non-uniqueness of BCJ numerators. The ambiguity with respect to total derivatives is naturally associated to the notion of twisted cohomology; see \cite{Mizera:2017cqs,Mizera:2019blq,Casali:2019ihm,Bhardwaj:2023vvm,Mazloumi:2024wys}.


\subsection{Field-theory limit of the chiral integrands}

In the field-theory limit ($\alpha'\to 0$), the integral over the modular parameter in the amplitude formula  \eqref{eq:Ac} is dominated by the region $\tau \to i\infty$, or equivalently $q:=e^{2\pi i\tau}\to0$; see e.g.~\cite{Green:1982sw,Tourkine:2013rda}. The torus then reduces to a nodal sphere, with the node giving rise to two additional marked points. This is easy to picture if we start from the Schottky parametrisation of the torus.\footnote{The Schottky parametrisation has long been used in string amplitudes, e.g.~\cite{DiVecchia:1988cy,Tseytlin:1990mv,Magnea:2004ai}. While it obscures the modular properties, it simplifies the field-theory degeneration.}
If $dz$ is the usual holomorphic differential, we perform the coordinate transformation
\begin{equation}
e^{2i\pi z} = \frac{(\sigma-\sigma_+)(\sigma_*-\sigma_-)}{(\sigma-\sigma_-)(\sigma_*-\sigma_+)}
\,,\qquad 
dz = \left(\frac{1}{\sigma - \sigma_+}-\frac{1}{\sigma - \sigma_-}\right)\frac{d\sigma}{2\pi i}\,.
\end{equation}
The torus is then parametrised like the Riemann sphere with two disks (around $\sigma_+$ and $\sigma_-$) excised; the boundaries of the disks are identified in a clockwise/counter-clockwise manner, which gives rise to the handle of the torus. In the degeneration limit $q\to 0$, the radius of the two disks vanishes as $\sqrt{|q|}$, so $\sigma_+$ and $\sigma_-$ become two additional marked points of the worldsheet.

We are interested in the behaviour of the chiral integrands in the limit $q\to0$. For the objects $ g^{(w)}_{ij}$ that appear in the ansatze, we have:
\begin{equation}
\label{eq:g1lim}
 g^{(1)}_{ij} \to 
 \pi i \,\frac{\sigma_{i+}\sigma_{j-}+\sigma_{i-}\sigma_{j+}}{\sigma_{ij}\sigma_{+-}}\,,
\end{equation}
where we use the notation\, $\sigma_{ab}=\sigma_a-\sigma_b$\,, with $a,b\in\{1,\cdots,n,+,-\}$; and, for $K\geq1$,
\begin{equation}
\label{eq:ghigherlim}
 g^{(2K)}_{ij} \to -2 \,\zeta(2K) \,, \qquad g^{(2K+1)}_{ij} \to 0  \,.
\end{equation}
In view of the cross-ratios in \eqref{eq:g1lim}, it is clear that our ansatze for the chiral integrands lead to $\SL(2,\C)$-invariant functions of the $n+2$ marked points.

Now we note the following identity:
\[
(2\pi i)^{n-1}\prod_{i=2}^n dz_i = \prod_{i=2}^n \left(d\sigma_i \, \frac{\sigma_{+-}}{\sigma_{i+}\sigma_{i-}} \right) = (-1)^n\frac{d^n\sigma}{\vol \SL(2,\C)}\; \sum_{\rho\in S_{n}}\frac{1}{\sigma_{+\rho(1)}\sigma_{\rho(2)\rho(3)}\cdots\sigma_{\rho(n)-}\sigma_{-+}} \,,
\]
where
\[
\frac{d^n\sigma}{\vol \SL(2,\C)} := (\sigma_{+-}\sigma_{-1}\sigma_{1+}) \prod_{i=2}^n d\sigma_i
\]
denotes the usual $\SL(2,\C)$ fixing. 

The chiral integrands for the superstring have a regular limit as $q\to 0$, such that there is a finite loop integrand in field theory. Then, the expression on the nodal sphere must admit an $\SL(2,\C)$-invariant form \cite{10.2969/jmsj/03920191}
\[
\label{eq:assumpq0}
(-1)^n(2\pi i)^3\, \lim_{q \to 0}  \,\cI_n \prod_{i=2}^n dz_i  \;=\; \frac{d^n\sigma}{\vol \SL(2,\C)}\; \sum_{\rho\in S_{n}}\frac{N(\rho(1),\rho(2),\cdots,\rho(n);\ell)}{\sigma_{+\rho(1)}\sigma_{\rho(2)\rho(3)}\cdots\sigma_{\rho(n)-}\sigma_{-+}} \,,
\]
where the numerators $N(\cdots;\ell)$ are Lorentz-invariant functions of only the external data and the loop momentum, i.e.~they are independent of the marked points.\footnote{We chose to normalise the numerators such that factors of $\pi$ are absent. We did this by placing, on the left-hand side, the factor $(2\pi i)^3=(2\pi i)^{n-1}/(2\pi i)^{n-4}$. The denominator $(2\pi i)^{n-4}$ cancels a factor arising naturally from $\cI_n$, as exemplified for $n=5$ in \eqref{eq:5pteg}, in view of \eqref{eq:g1lim}.} The expression above allows us to restrict our ansatze for $\cI_n$, by excluding certain contributions, e.g.~$g^{(1)}_{ij}g^{(1)}_{ji}$ at 6 points, or $g^{(1)}_{ij}g^{(1)}_{jk}g^{(1)}_{ki}$ at 7 points.

\subsection{Relation to the BCJ form of the field-theory limit}

We note that the expression \eqref{eq:assumpq0} is a natural one-loop generalisation of a tree-level relation. The tree-level relation is an equality, not a limit. Its right-hand side has neither $\ell_\mu$ nor the pair of `loop punctures' $\sigma_\pm$; instead, a pair of (arbitrarily chosen) `particle punctures' plays the analogous role, say $\sigma_1$ and $\sigma_n$. At tree level, the relation was observed in string theory \cite{Mafra:2011kj,Mafra:2011nv} and in the scattering equations (CHY) formalism of field theory \cite{Cachazo:2013iea}; see also \cite{Azevedo:2018dgo,He:2018pol,He:2019drm,Mafra:2022wml,Huang:2016tag,Schlotterer:2016cxa}. A crucial point is that the numerators $N$ are identified with the field-theory BCJ numerators. This is also what we expect at one loop in view of \eqref{eq:assumpq0}. To be more precise, we expect that the super-Yang-Mills theory (SYM) limit of the open superstring \eqref{eq:Ao} and the supergravity (SUGRA) limit of closed superstring \eqref{eq:Ac} lead, respectively, to the field-theory BCJ representation \cite{Bern:2010ue}
\begin{equation}
\label{eq:Aft}
 \mathcal{A}_n^{(1),\text{SYM}} = \int_{\mathbb R^D} d^D\ell\;\sum_{\gamma\in\Gamma_{n,3}^{(1)}} \frac{N_\gamma(\ell)\; c_\gamma}{D_\gamma} \,,
 \qquad 
 \mathcal{A}_n^{(1),\text{SUGRA}} = \int_{\mathbb R^D} d^D\ell\;\sum_{\gamma\in\Gamma_{n,3}^{(1)}} \frac{N_\gamma(\ell)\; \tilde N_\gamma(\ell)}{D_\gamma} \,.
\end{equation}
In these expressions, $\Gamma_{n,3}^{(1)}$ is the set of distinct $n$-point trivalent one-loop diagrams, $1/{D_\gamma}$ is the set of massless propagators for a given diagram, and $c_\gamma$ is the colour factor for a given diagram in SYM. The $n$-gon diagrams (where all $n$ external particles are directly attached to the loop) are `master diagrams', and their numerators are identified with those in \eqref{eq:assumpq0}, with $\ell$ flowing between $\rho(n)$ and $\rho(1)$. The numerators of all other trivalent diagrams are obtained from linear combinations of master numerators. For instance, an $(n-1)$-gon with massive corner 12 has the numerator
\[
\label{eq:N12}
N([1,2],3,\cdots,n;\ell):=N(1,2,3,\cdots,n;\ell)-N(2,1,3,\cdots,n;\ell)\,,
\]
which is represented pictorially in Figure 1.
\begin{figure}[t]
    \begin{align}
        \includegraphics[align=c,scale=0.45]{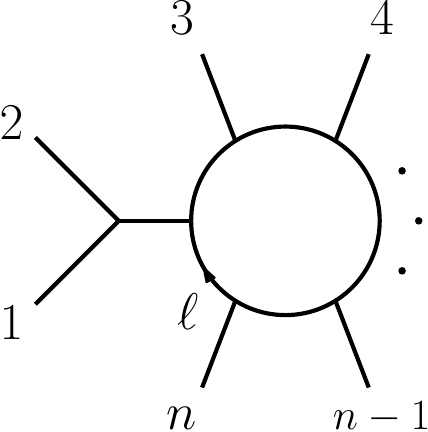} \quad = \quad
        \includegraphics[align=c,scale=0.45]{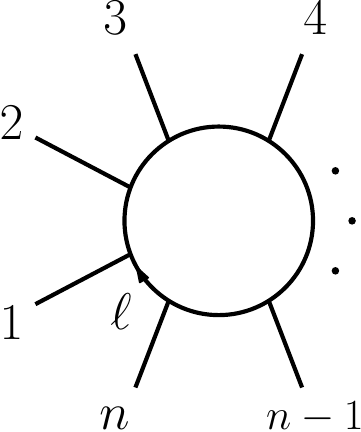} \quad - \quad
        \includegraphics[align=c,scale=0.45]{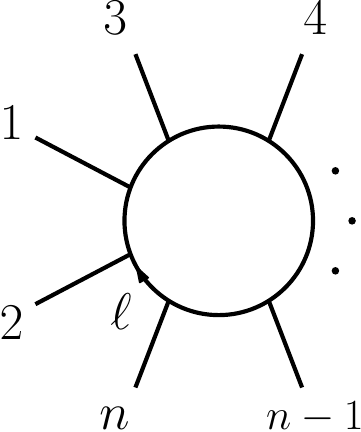} \nonumber
    \end{align}
\caption{Example of one-loop BCJ relation expressing a non-master numerator in terms of master numerators.}
\label{fig:N12}
\end{figure}  
The linear combinations are such that the numerators $N_\gamma$ of SYM satisfy the same Jacobi relations as the colour factors $c_\gamma$, hence the term colour-kinematics duality. Given the relation between unitarity constraints on the loop integrands in SYM and in SUGRA, if the SYM integrand can be expressed in such a colour-kinematics-dual way, then the SUGRA integrand is obtained for free as in \eqref{eq:Aft}: the colour factors of SYM are substituted by another copy (double copy) of SYM numerators $\tilde N_\gamma$. For instance, for NS-NS SUGRA external states with polarisations $\varepsilon_i^{\mu\nu}=\epsilon_i^\mu\tilde\epsilon_i^\nu$, the numerators $N_\gamma$ carry the polarisations $\epsilon_i^\mu$, while the numerators $\tilde N_\gamma$ carry the polarisations $\tilde\epsilon_i^\mu$. We suppressed coupling constants in \eqref{eq:Aft}. The algebraic properties of BCJ numerators, akin to those of the colour Lie algebra, reveal the existence of a `kinematic algebra' that has been widely investigated, e.g.~\cite{Mafra:2011kj,Monteiro:2011pc,Broedel:2012rc,Boels:2012ew,Boels:2013bi,Monteiro:2013rya,Cheung:2016prv,Du:2017kpo,Fu:2018hpu,Chen:2019ywi,Reiterer:2019dys,Frost:2020eoa,Borsten:2021hua,Chen:2021chy,Cheung:2021zvb,Brandhuber:2021bsf,Ben-Shahar:2021doh,Ben-Shahar:2021zww,Bonezzi:2022yuh,Brandhuber:2022enp,Borsten:2022vtg,Ben-Shahar:2022ixa,Borsten:2023ned,Bonezzi:2023lkx,Armstrong-Williams:2024icu,Chen:2024gkj,Bonezzi:2024fhd}. 

We emphasise that, in this paper, when we constrain the one-loop superstring amplitude in terms of the `field-theory limit', we mean the field-theory {\it loop integrand}, which can be obtained with modern unitarity techniques. In particular, we mean the loop integrand in the form \eqref{eq:Aft}. Obviously, we do not mean the field-theory {\it amplitude}, which is not defined in 10D due to the ultraviolet region of the loop integral. Our procedure will be to take a known representation \eqref{eq:Aft} and import the `master numerators' into the expression \eqref{eq:assumpq0}, in order to constrain the superstring ansatz.

The identification of the numerators in \eqref{eq:assumpq0} as BCJ numerators implies that they obey the symmetries expected from the corresponding $n$-gon diagrams. In particular, we note the following natural properties of the numerators: the cyclicity property,
\[
\label{eq:Ncyc}
N(1,2,3,\cdots,n-1,n;\ell) = N(2,3,\cdots,n-1,n,1;\ell+p_1)\,,
\]
and the reflection property,
\[
\label{eq:Nrefl}
N(1,2,3,\cdots,n-1,n;\ell) = (-1)^n N(n,n-1,\cdots,3,2,1;-\ell)\,.
\]
In addition, the BCJ numerators for the maximally-supersymmetric theories obey the following property: an $(m\leq n)$-gon numerator is a polynomial of order $(m-4)$ in the loop momentum $\ell_\mu$. Let us briefly discuss the consequences. The numerators appearing in \eqref{eq:assumpq0}, which are the $n$-gon numerators, have at most order $n-4$ in the loop momentum, which is consistent with the weight $n-4$ of $\cI_n$ associated to the table \eqref{eq:tablew}. Considering \eqref{eq:N12} and the fact that an $(n-1)$-gon BCJ numerator has at most order $n-5$ in the loop momentum, it follows that the $\ell^{n-4}$ part of the $n$-gon numerators, which is the leading order in $\ell$, is independent of the particle ordering. In fact, this property must cascade down to $(m< n)$-gons: the $\ell^{m-4}$ part of the $m$-gon numerators cannot depend on the order in which the $m$ trees are attached to the loop; this ensures that the $(m-1)$-gon numerators have no $\ell^{m-4}$ part. Cascading all the way down, we have that box (4-gon) numerators are independent of $\ell$ and of the order in which the four trees are attached to the loop, while triangle (3-gon) numerators simply vanish. We will provide examples later on that will illustrate these points. 

Finally, we note that there are consistency conditions on BCJ numerators coming from combinations of Jacobi-like relations that correspond to `trees going around the loop' \cite{Bjerrum-Bohr:2013iza}. We will see how these conditions map to monodromy constraints on $\cI_n$, e.g.~in \eqref{eq:N5monodromy} at 5 points. The consistency conditions establish the numerators in \eqref{eq:assumpq0} as standard BCJ numerators.\footnote{For the readers familiar with the one-loop ambitwistor string story, by standard we mean BCJ numerators on standard `quadratic' propagators, as opposed to the more exotic `linear' propagators.} We leave an explicit demonstration that
\eqref{eq:assumpq0} leads to the field-theory expressions \eqref{eq:Aft} for future work. On this note, there is a question about normalisation of the amplitudes. Ideally, we want the expressions \eqref{eq:Ao} and \eqref{eq:Ac} to directly match \eqref{eq:Aft} as $\alpha'\to0$. However, this requires an overall numerical factor in the superstring amplitudes (the overall $\alpha'$ factor is already correct). Using a result in \cite{Balli:2024wje}, we need to multiply the right-hand side of \eqref{eq:Ao} by $(2\pi i)^4$ for the open string, and the right-hand side of \eqref{eq:Ac} by $(-1)^n 2^8\pi^{8-n}$ for the open string. We will not dwell more on this point in the present paper.

In the following sections, we will see how our ansatze for the chiral integrands, subject to the degeneration limit \eqref{eq:assumpq0}, are completely fixed up to 7 points by the knowledge of the BCJ numerators $N$. Hence, the crucial ingredient $\cI_n$ of the superstring amplitude formula \eqref{eq:Ac} is fully determined by the integrand-level field-theory limit. This is remarkable, in view of the loss of information about the individual coefficients of the ansatze that is expected from the degeneration limit, in particular \eqref{eq:ghigherlim}. In short, the constraining power of monodromy invariance compensates for this loss of information up to 7 points, and possibly higher points.

\section{(Anti)symmetrisation notation}\label{sec:symmetrise}

In this brief section, we make a comment on notation. Due to the definition of BCJ numerators in terms of commutator-type combinations of master numerators, e.g.~\eqref{eq:N12}, we will try to  avoid confusion by using the following convention for (anti)symmetrisation:
\[
A_{ij} = \frac{A_{(ij)}+A_{[ij]}}{2}\,, \quad 
A_{(ij)} = A_{ij} + A_{ji} \,, \quad
A_{[ij]} = A_{ij} - A_{ji}\,.
\]
We note this in a separate section in order to (i) hopefully catch the reader's attention, and (ii) neatly line up the numbering of sections 4 to 7 with the 4- to 7-point chiral integrand construction.

\section{4-point chiral integrand}
\label{sec:4pt}

The chiral integrand $\cI_4$ has weight $n-4=0$, so it is just a constant on the torus, i.e.~a function only of the kinematic data, namely external momenta and polarisations (say for NS states):
\[
\cI_4 = C_4(\epsilon_i,p_i)\,.
\]
Using \eqref{eq:assumpq0}, we read off the numerators $N$, which are all identified with
\[
\label{eq:N4C4}
N(1234;\ell) = C_4(\epsilon_i,p_i)\,.
\]
We will review in section~\ref{sec:4ptexplicit} the long-known explicit expression for $C_4(\epsilon_i,p_i)$.
Notice that there is no dependence on the loop momentum. Moreover, $\cI_4$ does not depend on any particle ordering, so neither does $C_4(\epsilon_i,p_i)$. Hence, the $4!$ numerators $N(\rho)$, with $\rho\in S_4$, are identical. Interpreting these as BCJ numerators, we conclude immediately that the numerators of `triangle' diagrams vanish, e.g.~if the massive corner is 12, we have $N([1,2]34)=N(1234)-N(2134)=0$\,. These features match the expected properties of BCJ numerators discussed below \eqref{eq:Nrefl}.

We could, of course, have started from the knowledge of the BCJ numerators, and employed \eqref{eq:N4C4} to identify the kinematic function $C_4(\epsilon_i,p_i)$ corresponding to $\cI_4$. This is the perspective that we are taking in this paper, although the 4-point example here is somewhat trivial.

\section{5-point chiral integrand}
\label{sec:5pt}

The 5-point ansatz was already mentioned in \eqref{eq:5pteg},
\[
\label{eq:I5ansatz}
\boxed{
\cI_5 = 2\pi i\;C_5^\mu \,\ell_\mu + \sum_{\substack{i,j\\
                  i<j}}\, C_{5,ij}\, g^{(1)}_{ij}\,.
                  }
\]
Notice that $\cI_5$ does not depend on any particle ordering, so the coefficient $C_5^\mu$ shares this property. Now, we use \eqref{eq:assumpq0} to read off
\[
N(12345;\ell) = C_5^\mu \,\ell_\mu - \frac1{2} \,\sum_{\substack{i,j\\
                  i<j}}\, C_{5,ij}\,.
\label{eq:N5C}
\]
All other numerators appearing in \eqref{eq:assumpq0} are obtained by relabelings of the external particles.\footnote{Due to the antisymmetry $g^{(1)}_{ij}=-g^{(1)}_{ji}$, it is natural to define  $C_{5,ji}=-C_{5,ij}$. In this way, we have \,$$\sum_{\substack{i,j\\
                  i<j}}\, C_{5,ij}\, g^{(1)}_{ij}=\frac1{2}\sum_{\substack{i,j\\
                  i\neq j}}\, C_{5,ij}\, g^{(1)}_{ij}\,.$$
                  This natural extension of coefficients (here and in more complicated later examples) will be useful when considering different particle orderings. 
                  \label{footnote:ji}}

We expressed the numerators in terms of the kinematic coefficients of the ansatz \eqref{eq:I5ansatz}. What may not be immediately obvious, although it is straightforward to check, is that we can invert the relation \eqref{eq:N5C} to express the coefficients in terms of the numerators:
\[
\boxed{
C_5^\mu = N(\cdots;\ell)|_{\ell_\mu}\,, \qquad C_{5,ij} = - N(\cdots[i,j]\cdots;\ell)=:-N([i,j])\,.
}
\label{eq:C5N}
\]
Therefore, if we start from the knowledge of the BCJ numerators, we can construct $\cI_5$. We extract $C_5^\mu$ as the coefficient of the term linear in loop momentum. Following our discussion below \eqref{eq:Nrefl}, this coefficient must be independent of the particle ordering of the pentagon numerators, and this is the reason why we need not specify a particular ordering in $N(\cdots;\ell)|_{\ell_\mu}$. Moreover, the coefficients $C_{5,ij}$ are obtained from the difference between two pentagon numerators. Here, the $\ell$-dependence drops off, and the ordering of the other three particles (not $i,j$) is irrelevant, again due to general properties of BCJ numerators.\footnote{Note, for instance, that $N([1,2][3,4]5;\ell)$ and $N([[1,2],3],4,5;\ell)$ are triangle (3-gon) BCJ numerators, and hence vanish.}

Let us address now the behaviour of $\cI_5$ under a monodromy transformation \eqref{eq:monodromy}. Invariance requires
\[
 0=\cI_5|^{\ell\mapsto\ell-p_i}_{z_i\mapsto z_i+\tau} - \cI_5= -2\pi i \Bigg(C_5^\mu \,{p_i}_\mu + \sum_{j\neq i}\, C_{5,ij}\Bigg)  \,.
 \label{eq:5ptmonodromy}
\]
Since we can construct $\cI_5$ from the knowledge of 5-point BCJ numerators, the numerators had better know about this identity. Indeed, there is an equivalent `BCJ monodromy' that constrains the numerators, as observed e.g.~in \cite{Bjerrum-Bohr:2013iza}. To check this, let us take particle $i=1$ around the loop,
\begin{align}
C_5^\mu \,{p_1}_\mu \,&= N(12345;\ell)- N(12345;\ell-p_1) = N(12345;\ell)- N(23451;\ell) =
\nonumber \\
&= N(12345;\ell) - N(21345;\ell) + N(21345;\ell) - N(23145;\ell) + N(23145;\ell) - N(23415;\ell) \nonumber \\ &\qquad + N(23415;\ell) - N(23451;\ell) =
\nonumber \\
&= N([1,2]345;\ell) + N(2[1,3]45;\ell) +N(23[1,4]5;\ell) + N(234[1,5];\ell)=
\nonumber \\
&= \sum_{j>1} N([1,j]) = - \sum_{j>1}\, C_{5,1j}\,,
\label{eq:N5monodromy}
\end{align}
where we used the property \eqref{eq:Ncyc} in the second equality.

Before proceeding, let us note that one could argue {\it a priori} that the BCJ numerators take the form  \eqref{eq:N5C}, with identifications \eqref{eq:C5N}. We have already mentioned that the piece linear in $\ell$ must be invariant under permutations of the particle ordering, because box numerators $N([i,j])$ do not depend on $\ell$. As for the piece independent of $\ell$, it cannot give rise to triangle numerators, and notice also that it cannot have a permutation symmetric part, due to the reflection property \eqref{eq:Nrefl}. This leads to \eqref{eq:N5C}-\eqref{eq:C5N}.

\section{6-point chiral integrand}
\label{sec:6pt}

The 6-point ansatz is
\begin{align}
\cI_6 = \; & (2\pi i)^2\, C_{6}^{\mu\nu} \,\ell_\mu\ell_\nu + 2\pi i \sum_{\substack{i,j\\
                  i<j}}\, C_{6,ij}^\mu \, \ell_\mu\,  g^{(1)}_{ij}  \;+ \sum_{\substack{i,j,k,l \text{\,dist}\\ i<j,k<l,i<k}}\, C_{6,ij,kl}\,  g^{(1)}_{ij}\,  g^{(1)}_{kl} \nonumber\\
                  & \;\;  + \sum_{\substack{i,j,k \text{\,dist}\\ i<k}}\, C_{6,ijk}\,  g^{(1)}_{ij}\,  g^{(1)}_{kj} \;+\; \sum_{\substack{i,j\\
                  i<j}}\, \tilde C_{6,ij}\,  g^{(2)}_{ij} \,,
                  \label{eq:I6initial}
\end{align}
where, `dist' in the sums means `all distinct'.
According to our rules, we excluded terms of the type $g^{(1)}_{ij}g^{(1)}_{ij}$. There is an ambiguity in the ansatz coefficients, due to the Fay relations
\[
g^{(1)}_{ij} g^{(1)}_{jk} + g^{(2)}_{ik} + \text{cyc}(i,j,k)=0\,.
\]
If we add the vanishing expression\footnote{We recall that, in our convention, $C_{i[jk]} = C_{ijk}-C_{ikj}$. Similarly to footnote \ref{footnote:ji}, one can extend the definition of $C_{ijk}$ to $k>i$ using $C_{ijk}=C_{kji}$, in accordance with the symmetry of $g^{(1)}_{ij} g^{(1)}_{kj}$ in the corresponding term of \eqref{eq:I6initial}.\label{footnote:jik}}
\[
\label{eq:6ptfayzero}
\frac1{18} \sum_{\substack{i,j,k \text{\,dist}}} (C_{ijk}+C_{jki}+C_{kij}) \left(g^{(1)}_{ij} g^{(1)}_{jk} + g^{(2)}_{ik} + \text{cyc}(i,j,k)\right)\,,
\]
after some algebra we obtain
\begin{empheq}[box=\widefbox]{align}
\cI_6 =\; & (2\pi i)^2\, C_{6}^{\mu\nu} \,\ell_\mu\ell_\nu \;+ 2\pi i \sum_{\substack{i,j\\
                  i<j}}\, C_{6,ij}^\mu \, \ell_\mu\,  g^{(1)}_{ij} \; + \sum_{\substack{i,j,k,l \text{\,dist}\\ i<j,k<l,i<k}}\, C_{6,ij,kl}\,  g^{(1)}_{ij}\,  g^{(1)}_{kl} \nonumber\\
                  & \;\;  +\; \frac1{3}\sum_{\substack{i,j,k \text{\,dist}\\ i<k}}\, (C_{6,i[jk]}+C_{6,k[ji]})\,  g^{(1)}_{ij}\,  g^{(1)}_{kj}  \;+\; \sum_{\substack{i,j\\
                  i<j}}\, \tilde C'_{6,ij}\,  g^{(2)}_{ij} \,,
                  \label{eq:I6fay}
                  \end{empheq}
where
\[
\label{eq:Ctildeprime}
\tilde C'_{6,ij} = \tilde C_{6,ij} +\frac1{3} \sum_{k\neq i,j} (C_{k(ij)}+C_{ikj})\,.
\]
The latter expression will not be important, because we will fix $\tilde C'_{6,ij}$ directly from the monodromy relations in terms of the other coefficients in the ansatz \eqref{eq:I6fay}. What we gain by working with the expression \eqref{eq:I6fay} -- as opposed to \eqref{eq:I6initial} -- is that, unlike $C_{6,ijk}$, the combination $C_{6,i[jk]}$ will be unambiguously identified from the field-theory limit.

In the degeneration limit associated to field theory, we read off the master BCJ numerators from \eqref{eq:I6fay} using \eqref{eq:assumpq0}:\footnote{We employ here the natural Mathematica notation that $\text{If}[i<j<k,1,0]$ is $1$ if $i<j<k$, and $0$ otherwise.} 
\begin{align}
N(123456;\ell) 
=\; & C_{6}^{\mu\nu} \,\ell_\mu\ell_\nu - \frac1{2} \,\sum_{\substack{i,j\\
                  i<j}}\, C_{6,ij}^\mu\,\ell_\mu +\frac1{4} \sum_{\substack{i,j,k,l \text{\,dist}\\ i<j,k<l,i<k}}\, C_{6,ij,kl} \nonumber \\
& +  \frac1{12}\sum_{\substack{i,j,k \text{\,dist}\\ i<k}}\, (-1)^{\text{If}[i<j<k,1,0]}\,(C_{6,i[jk]}+C_{6,k[ji]}) + \frac1{12} \,\sum_{\substack{i,j\\
                  i<j}}  \tilde C'_{6,ij}\,.
                  \label{eq:N6}
\end{align}
We can now invert this relation to read off the following coefficients of our ansatz \eqref{eq:I6fay}:

\begin{empheq}[box=\widefbox]{align}
& C_6^{\mu\nu} = N(\cdots;\ell)|_{\ell_\mu\ell_\nu}\,, \qquad C_{6,ij}^\mu = - N(\cdots[i,j]\cdots;\ell)|_{\ell_\mu}\,, \nonumber \\ &C_{6,ij,kl}  =  N(\cdots[i,j]\cdots[k,l]\cdots;\ell)\,, \qquad
 C_{6,i[jk]} =- N(\cdots[i,[j,k]]\cdots;\ell)\,.
 \label{eq:C6N6}
  \end{empheq}
The ellipses mean that the particle ordering is irrelevant, with the exception of what is explicitly prescribed. In the coefficients of the second line, the loop-momentum dependence drops out from the combination of numerators. We are only missing an expression for the coefficients $\tilde C'_{6,ij}$ in terms of the numerators. These coefficients cannot be individually discerned in the $q\to0$ limit of \eqref{eq:I6fay}, because in this limit $g^{(2)}_{ij}\to -2 \,\zeta(2)$, independently of the pair $\{i,j\}$. However, as we will now see, the monodromy constraint comes to the rescue.

Using $\mathcal{I}_6$ given in equation \eqref{eq:I6fay}, the monodromy relation obtained at 6 points is
\[
\label{eq:DiI6}
0\;=\; \cI_6|^{\ell\mapsto\ell-p_i}_{z_i\mapsto z_i+\tau} - \cI_6 \;=\; -2\pi i\Bigg( \pi i\left(2\,\ell_\mu-p_{i\mu}\right)  M^\mu_i+\sum_{j\neq i}M_{ij}\left( g^{(1)}_{ij}-\pi i \right) +\sum_{\substack{j,k\neq i\\ j<k}} M_{ijk}\,g^{(1)}_{jk}\Bigg) 
\,,
\]
where the coefficients $M^\mu_i$, $M_{ij}$ and $M_{ijk}$ depend only on the external particle data. These coefficients must vanish separately, because the monodromy relations hold for arbitrary moduli, including the loop momentum. Explicitly,
\begin{align}
\label{eq:Mmui6}
	M^\mu_i &= 2\,p_{i\nu}C_6^{\mu\nu} +\sum_{j\neq i} C^\mu_{6,ij}\, ,\\
	M_{ij} &= p_{i\mu} C^\mu_{6,ij} +\, \frac1{3}\sum_{k\neq i,j}\, (C_{6,j[ik]}+C_{6,k[ij]}) \;+\tilde{C}'_{6,ij} \,.  \label{eq:Mmui6ij}\\
	M_{ijk} &= p_{i\mu} C^\mu_{6,jk} + \sum_{l\neq i,j,k} C_{6,il,jk} \;- C_{i\left[jk\right]} \,.
 \label{eq:M6ijk}
\end{align}
Regarding $M_{ij}$, it is useful to consider separately
\begin{align}
\label{eq:M6ijanti}
 	M_{[ij]} &= (p_{i\mu}+p_{j\mu}) C^\mu_{6,ij} + \sum_{k\neq i,j} C_{6,k[ij]} \,,  \\
	M_{(ij)} &= (p_{i\mu}-p_{j\mu}) C^\mu_{6,ij} + \frac1{3}\sum_{k\neq i,j}\, (C_{6,j[ik]}+C_{6,i[jk]})\; + 2\, \tilde{C}'_{6,ij} \,.
\end{align}
The monodromy relation $M_{(ij)}=0$ implies
\begin{empheq}[box=\widefbox]{align}
\tilde{C}'_{6,ij} = -\frac1{2} \Big(
(p_{i\mu}-p_{j\mu}) C^\mu_{6,ij} + \frac1{3}\sum_{k\neq i,j}\, (C_{6,j[ik]}+C_{6,i[jk]})
\Big) \,.
\label{eq:C6N6tilde}
  \end{empheq}
This fixes the coefficients $\tilde{C}'_{6,ij}$ in terms of those coefficients that we already know how to extract from the BCJ numerators via \eqref{eq:C6N6}.
Hence, all the coefficients in the chiral integrand \eqref{eq:I6fay} are now determined in terms of the field-theory limit.

Similarly to the 5-point case \eqref{eq:N5monodromy}, the monodromy relations $M^\mu_i=0$, $M_{[ij]}=0$ and $M_{ijk}=0$ correspond to `BCJ monodromies' where one particle or a multi-particle tree is taken around the loop. The case of $M^\mu_i=0$ is a 6-point analogue of \eqref{eq:N5monodromy}, where particle $i$ goes around the loop, but now we pick up the contribution linear in $\ell$; and the piece independent of $\ell$ turns out to not add any additional constraint. The case of $M_{[ij]}=0$ is also an analogue of \eqref{eq:N5monodromy}, but one where the bi-particle $[i,j]$ goes around the loop. Finally, the case of $M_{ijk}=0$ is one where particle $i$ goes around a loop which possesses a massive corner $[j,k]$. We present details of these BCJ monodromies in Appendix~\ref{app:6ptBCJmonodromies}. The upshot is that these monodromy relations are already known constraints among BCJ numerators. The only monodromy relation that is not of this type, namely $M_{(ij)}=0$, plays instead the role of determining the remaining coefficients of the superstring ansatz via \eqref{eq:C6N6tilde}.

Before proceeding, let us make two remarks. The first remark is that, while finding the `perfect zero' \eqref{eq:6ptfayzero} was convenient, it is not essential. It allowed us to write $\cI_6$ in \eqref{eq:I6fay} uniquely in terms of BCJ numerators via the map (\ref{eq:C6N6}, \ref{eq:C6N6tilde}), and also allowed us later on to identify monodromy constraints with `BCJ monodromies'. Had we not bothered to find the `perfect zero', the coefficients of $\cI_6$ in \eqref{eq:I6initial} would not have been uniquely determined in terms of the numerators, but the remaining ambiguity would drop out fully from $\cI_6$ due to the Fay relations.

The second remark is about the BCJ structure of the amplitudes: the superstring integrand gives rise to BCJ numerators that are not of the most general type one could encounter at $n=6$ within our $\ell$-power-counting rules. We saw that we can write a one-to-one map between the chiral integrand \eqref{eq:I6fay} and the BCJ master numerators \eqref{eq:N6}. For instance, starting with the master numerators, one can obtain the chiral integral using (\ref{eq:C6N6}, \ref{eq:C6N6tilde}). Now, suppose we would add an additional $\ell$-independent permutation-symmetric term to the numerators \eqref{eq:N6}. On the one hand, this term would completely drop out in the map (\ref{eq:C6N6}, \ref{eq:C6N6tilde}), which involves only differences between numerators in what concerns the coefficients of the $\ell^0$ part of the chiral integrand. Hence, this term could not be recovered from \eqref{eq:assumpq0}. Therefore, the $\ell$-independent permutation-symmetric part of the master numerators obtained from the superstring is not independent of the other parts of the numerators. On the other hand, adding a new term to the master (hexagon) numerators that is $\ell$-independent and permutation-symmetric would not alter their BCJ properties: the cyclic and reflection symmetries would still hold, as would the usual BCJ relations, in which only differences between master numerators appear, e.g.~\eqref{eq:N12}. We conclude that the numerators are {\it a priori} more constrained by their superstring origins than what the BCJ relations alone imply.

\section{7-point chiral integrand}
\label{sec:7pt}

The 7-point ansatz is
\begin{align}
    \cI_7=\, & (2\pi i)^3\, C_7^{\mu \nu \rho} \ell_\mu \ell_\nu \ell_\rho
    + (2\pi i)^2 \sum_{\substack{i, j \\ i<j}} C_{7, i j}^{\mu \nu}\, \ell_\mu \ell_\nu\, g_{i j}^{(1)}
    + 2\pi i\!\sum_{\substack{i, j, k, l \text { dist } \\ i<j, k<l, i<k}} \!C_{7, i j, k l}^\mu\, \ell_\mu \,g_{i j}^{(1)} g_{k l}^{(1)} \nonumber \\
    & +\frac{2\pi i}{3} \sum_{\substack{i, j, k \text { dist } \\
    i<k}}\left(C_{7, i[jk]}^\mu + C_{7, k[j i]}^\mu\right) \ell_\mu \, g_{i j}^{(1)} g_{k j}^{(1)}
    + 2\pi i\sum_{\substack{i, j \\ i<j}} \tilde{C}_{7, i j}^{\prime \mu}\, \ell_\mu \,g_{i j}^{(2)} \nonumber \\
    & +\sum_{\substack{i, j, k, l, m, n \text { dist } \\ i<j, k<l, m<n, i<k<m}} C_{7, i j, k l, m n}\, g_{i j}^{(1)} g_{k l}^{(1)} g_{m n}^{(1)}
    + \sum_{\substack{i, j, k, l, m \text { dist } \nonumber \\ i<k, l<m}} C_{7, i j k, l m}\, g_{i j}^{(1)} g_{k j}^{(1)} g_{l m}^{(1)} \\
    & +\sum_{\substack{i, j, k, l \text { dist } \\ i<l}} C_{7, i j k l}\, g_{i j}^{(1)} g_{j k}^{(1)} g_{k l}^{(1)}
    + \sum_{\substack{i, j, k, l \text { dist } \nonumber \\ i <j, k<l}} \tilde{C}_{7, i j, k l}\, g_{i j}^{(1)} g_{k l}^{(2)} \\
    & +\sum_{i, j, k \text { dist }} \tilde{C}_{7, i j k}\, g_{i j}^{(1)} g_{k i}^{(2)}
    + \sum_{\substack{i, j \\ i<j}} \tilde{C}_{7, i j} \,g_{i j}^{(3)} .
    \label{eq:I7}
\end{align}
The first two lines are entirely analogous to the 6-point chiral integrand \eqref{eq:I6fay}, with an additional factor $2\pi i\ell_\mu$ adjusting the weight to what is required at 7 points. So the $\ell$-dependent part of $\cI_7$ is clear. The $\ell$-independent part, in the last three lines, requires more thinking. Similarly to the 6-point case, we have excluded terms with cyclic puncture indices, such as $g_{i j}^{(1)} g_{ij}^{(2)}$, $g_{i j}^{(1)} g_{j k}^{(1)} g_{k i}^{(1)}$, and powers of $g_{ij}^{(1)}$ with the same $(i,j)$. In addition, we have to consider the Fay relations up to this order, namely
\begin{equation}\label{eq:Fay2}
   g^{(2)}_{ik} + g^{(2)}_{ji} + g^{(2)}_{kj} +  g^{(1)}_{ij} g^{(1)}_{jk} + g^{(1)}_{jk} g^{(1)}_{ki} + g^{(1)}_{ki} g^{(1)}_{ij}
    = 0\,,
\end{equation}
which was relevant already at 6 points, and now also
\begin{equation}\label{eq:Fay3}
    g_{i k}^{(3)}+g_{k j}^{(3)}-2 g_{j i}^{(3)}+g_{i j}^{(2)} g_{j k}^{(1)}+g_{k i}^{(1)} g_{i j}^{(2)}-g_{j k}^{(2)} g_{k i}^{(1)}-g_{j k}^{(1)} g_{k i}^{(2)}=0\,.
\end{equation}
One immediate consequence is that we were allowed to exclude in \eqref{eq:I7} terms of the type $g^{(1)}_{ij} g^{(1)}_{ik} g^{(1)}_{il}$, because the weight-2 Fay relations \eqref{eq:Fay2} permit their replacement by terms already in the ansatz. Now, we recall that, at 6 points, we identified the zero \eqref{eq:6ptfayzero} to be added to the chiral integrand $\cI_6$, going from \eqref{eq:I6initial} to \eqref{eq:I6fay}, in order to write $\cI_6$ manifestly with coefficients uniquely determined from BCJ numerators. Note that this vanishing contribution does not alter the chiral integrand, and does not alter the BCJ numerators obtained via the degeneration limit \eqref{eq:assumpq0}. We implicitly did the same thing at 7 points for the $\ell$-dependent part of \eqref{eq:I7}. For the same reason, but now for the $\ell$-independent part, we will add the following vanishing expression:
\begin{equation}
    0 = X_0 + X_1 + X_2 \,,
\end{equation}
with\footnote{Similarly to the 6-point case, as noted in footnotes \ref{footnote:ji} and \ref{footnote:jik}, we can extend the coefficients appearing in \eqref{eq:I7} to the region outside the summation according to the (anti)symmetry of the associated $g$-functions:
\begin{align}
       C^{\mu}_{7,ijk} = C^{\mu}_{7,kji} \,, \quad & \quad C_{7,ijk,lm}^{\mu\nu} = C_{7,kji,lm}^{\mu\nu} = - C_{7,ijk,ml}^{\mu\nu} = - C_{7,kji,ml}^{\mu\nu} \,, \nonumber \\
        C_{7,ijkl} = - C_{7,lkji} \,, \quad & \quad \tilde{C}_{7,ij,kl} = -\tilde{C}_{7,ji,kl} = \tilde{C}_{7,ij,lk} = - \tilde{C}_{7,ji,lk} \,.
\end{align}
}
\begin{equation}
    X_0 = \frac{1}{18} \sum_{\substack{i,j,k,l,m\text{ dist } \\ l<m}} (C_{7,ijk,lm} + C_{7,jki,lm} + C_{7,kij,lm}) \, g_{lm}^{(1)} \, \left( g_{ij}^{(1)} g_{jk}^{(1)} + g_{ik}^{(2)} + \text{cyc}(i,j,k) \right) \,,
\end{equation}
\begin{equation}
    \begin{aligned}
        X_1 = \frac{1}{4} \sum_{\substack{i,j,k,l \text{\,dist} \\ i<l}} C_{7,ijkl} \Big\lbrack 
        & \left( g_{il}^{(1)} - 2 g_{jl}^{(1)} - 2 g_{kl}^{(1)} \right) 
        \left( g_{ij}^{(2)} + g_{ik}^{(2)} + g_{jk}^{(2)} - g_{ij}^{(1)} g_{ik}^{(1)} + g_{ij}^{(1)} g_{jk}^{(1)} - g_{ik}^{(1)} g_{jk}^{(1)}\right) \\
        & + \left( g_{il}^{(1)} - 2 g_{ij}^{(1)} - 2 g_{ik}^{(1)} \right) 
        \left( g_{jk}^{(2)} + g_{jl}^{(2)} + g_{kl}^{(2)} - g_{jk}^{(1)} g_{jl}^{(1)} + g_{jk}^{(1)} g_{kl}^{(1)} - g_{jl}^{(1)} g_{kl}^{(1)} \right) \\
        & - g_{ij}^{(1)} 
        \left( g_{ik}^{(2)} + g_{il}^{(2)} + g_{kl}^{(2)} - g_{ik}^{(1)} g_{il}^{(1)} + g_{ik}^{(1)} g_{kl}^{(1)} - g_{il}^{(1)} g_{kl}^{(1)} \right) \\
        & - g_{kl}^{(1)} 
        \left( g_{ij}^{(2)} + g_{il}^{(2)} + g_{jl}^{(2)} - g_{ij}^{(1)} g_{il}^{(1)} + g_{ij}^{(1)} g_{jl}^{(1)} - g_{il}^{(1)} g_{jl}^{(1)} \right) 
        \Big\rbrack \,,
    \end{aligned}
\end{equation}
\begin{equation}
    \begin{aligned}
        X_2 = \frac{1}{6} \sum_{i,j,k \text{\, dist}} \tilde{C}_{7,ijk} \Big\lbrack 
        & \left( g_{ji}^{(3)} + g_{ik}^{(3)} - 2g_{kj}^{(3)} + g_{jk}^{(2)} g_{ki}^{(1)} + g_{ij}^{(1)} g_{jk}^{(2)} - g_{ki}^{(2)} g_{ij}^{(1)} - g_{ki}^{(1)} g_{ij}^{(2)} \right) \\
        & - \left( g_{kj}^{(3)} + g_{ji}^{(3)} - 2g_{ik}^{(3)} + g_{ki}^{(2)} g_{ij}^{(1)} + g_{jk}^{(1)} g_{ki}^{(2)} - g_{ij}^{(2)} g_{jk}^{(1)} - g_{ij}^{(1)} g_{jk}^{(2)} \right)
        \Big\rbrack \,.
    \end{aligned}
\end{equation}
After some algebra, we obtain
\begin{empheq}[box=\widefbox]{align}
    \cI_7=\, & (2\pi i)^3\, C_7^{\mu \nu \rho} \ell_\mu \ell_\nu \ell_\rho
    + (2\pi i)^2 \sum_{\substack{i, j \\ i<j}} C_{7, i j}^{\mu \nu}\, \ell_\mu \ell_\nu\, g_{i j}^{(1)}
    + 2\pi i\!\sum_{\substack{i, j, k, l \text { dist } \\ i<j, k<l, i<k}} \!C_{7, i j, k l}^\mu\, \ell_\mu \,g_{i j}^{(1)} g_{k l}^{(1)} \nonumber \\
    & +\frac{2\pi i}{3} \sum_{\substack{i, j, k \text { dist } \\
    i<k}}\left(C_{7, i[jk]}^\mu + C_{7, k[j i]}^\mu\right) \ell_\mu \, g_{i j}^{(1)} g_{k j}^{(1)}
    + 2\pi i\sum_{\substack{i, j \\ i<j}} \tilde{C}_{7, i j}^{\prime \mu}\, \ell_\mu \,g_{i j}^{(2)} \nonumber \\
    & +\sum_{\substack{i, j, k, l, m, n \text { dist } \\ i<j, k<l, m<n, i<k<m}} C_{7, i j, k l, m n}\, g_{i j}^{(1)} g_{k l}^{(1)} g_{m n}^{(1)}
    \nonumber \\
    &
    + \frac{1}{3}\sum_{\substack{i, j, k, l, m \text { dist } \nonumber \\ i<k, l<m}} \left( C_{7, i [j k], l m} + C_{7, k [j i], l m} \right)\, g_{i j}^{(1)} g_{k j}^{(1)} g_{l m}^{(1)} \\
    & +\sum_{\substack{i, j, k, l \text { dist } \\ i<l}} C'_{7, i j k l}\, g_{i j}^{(1)} g_{j k}^{(1)} g_{k l}^{(1)}
    + \sum_{\substack{i, j, k, l \text { dist } \nonumber \\ i <j, k<l}} \tilde{C}'_{7, i j, k l}\, g_{i j}^{(1)} g_{k l}^{(2)} \\
    & +\sum_{i, j, k \text { dist }} \tilde{C}'_{7, i j k}\, g_{i j}^{(1)} g_{k i}^{(2)}
    + \sum_{\substack{i, j \\ i<j}} \tilde{C}'_{7, i j} \,g_{i j}^{(3)} ,
    \label{eq:I7fay}
\end{empheq}
where\footnote{Following our notation for (anti)symmetrisation of two points, defined in section~\ref{sec:symmetrise}, we use the following convention for symmetrisation of three points: \, $A_{(ijk)} = A_{(ij)k} + A_{(jk)i} + A_{(ki)j}$ \,.
}
\begin{align}
    C'_{7,ijkl} =&\, \frac{1}{8}\, \Big( 4 C_{7,ijkl}-C_{7,ijlk}-4 C_{7,ikjl}+C_{7,iklj}-2 C_{7,iljk}+C_{7,ilkj}-C_{7,jikl} \\
        & \qquad + 2 C_{7,jilk}-2 C_{7,jkil}+C_{7,kijl}+C_{7,kjil} \Big) \,, \nonumber\\
    \tilde{C}'_{7,ijk} =&\, \frac{1}{2}\, \tilde{C}_{7,[ij]k} + \frac{1}{6} \tilde{C}_{7,(ijk)} \,, \label{eq:newC7ijk}\\
    \tilde{C}'_{7,ij,kl} =&\, \tilde{C}_{7,ij,kl} + \frac{1}{3} \sum_{m \neq i,j,k,l} \left( C_{7,ij,klm} + C_{7,ij,lmk} + C_{7,ij,mkl} \right) \,, \\
    \tilde{C}'_{7,ij} =&\, \tilde{C}_{7,ij} + \frac{1}{2} \sum_{k \neq i,j} \left( \tilde{C}_{7,(ik)j} - \tilde{C}_{7,(jk)i} \right) \,.
\end{align}
The latter relations will not be important, because we want to work only with the form \eqref{eq:I7fay} of the ansatz from now on. Similarly to the 6-point case, this will enable us to write the coefficients $C'_{7,ijkl}$ uniquely in terms of the BCJ numerators, and later we will use the monodromy relations to determine also the `tilded' coefficients $\tilde{C}'_{7;ij,kl}$, $\tilde{C}'_{7,ijk}$ and $\tilde{C}'_{7,ij}$ in terms of BCJ numerators.

In the degeneration limit \eqref{eq:assumpq0}, we can read off the master BCJ numerators from \eqref{eq:I7fay}:
\begin{equation}
    \begin{aligned}
        N(1234567;\ell) =\, &  C_7^{\mu \nu \rho} \ell_\mu \ell_\nu \ell_\rho
        - \frac{1}{2} \sum_{\substack{i, j \\ i<j}} C_{7, i j}^{\mu \nu}\, \ell_\mu \ell_\nu 
        + \frac{1}{4} \! \sum_{\substack{i, j, k, l \text { dist } \\ i<j, k<l, i<k}} \!C_{7, i j, k l}^\mu\, \ell_\mu \\
        & +\frac{1}{12} \sum_{\substack{i, j, k \text { dist } \\ i<k}} (-1)^{\text{If[$i<j<k$,1,0]}} \left(C_{7, i[jk]}^\mu + C_{7, k[j i]}^\mu\right) \ell_\mu 
        + \frac{1}{12} \sum_{\substack{i, j \\ i<j}} \tilde{C}_{7, i j}^{\prime \mu}\, \ell_\mu \\
        & - \frac{1}{8} \sum_{\substack{i, j, k, l, m, n \text { dist } \\ i<j, k<l, m<n, i<k<m}} \!\!\! C_{7, i j, k l, m n}\\
        &
        - \frac{1}{24} \sum_{\substack{i, j, k, l, m \text { dist } \\ i<k, l<m}} (-1)^{\text{If[$i<j<k$,1,0]}} \left( C_{7, i [j k], l m} + C_{7, k [j i], l m} \right) \\
        & +\frac{1}{8} \sum_{\substack{i, j, k, l \text { dist } \\ i<l}} \text{sign}\big\lbrack (i-j) (j-k) (k-l) \big\rbrack C'_{7, i j k l}
        - \frac{1}{12} \sum_{\substack{i, j, k, l \text { dist } \\ i <j, k<l}} \tilde{C}'_{7, i j, k l} \\
        & - \frac{1}{12} \sum_{\substack{i,j,k \text{ dist } \\ i<j}} \tilde{C}'_{7, i j k} \,.
    \end{aligned}
\end{equation}
Partly inverting this relation, we can express the following coefficients in terms of numerators:
\begin{empheq}[box=\widefbox]{equation}
    \begin{aligned}
        & C_7^{\mu\nu\rho} = N(\cdots;\ell)|_{\ell_\mu \ell_\nu \ell_\rho} \,, \qquad C_{7,ij}^{\mu\nu} = - N(\cdots[i,j]\cdots;\ell)|_{\ell_\mu \ell_\nu} \,, \\ 
        & C_{7,ij,kl}^{\mu}  =  N(\cdots[i,j]\cdots[k,l]\cdots;\ell)|_{\ell_\mu} \,, \qquad
        C_{7,i[jk]}^{\mu} = - N(\cdots[i,[j,k]]\cdots;\ell)|_{\ell_\mu} \,, \\
        & C_{7,ij,kl,mn} = - N(\cdots [i,j] \cdots [k,l] \cdots [m,n]; \ell) \,, \quad \\
        & C_{7,i[jk],lm} = N(\cdots [i,[j,k]] \cdots [l,m] \cdots; \ell) \,, \\
        & C'_{7,ijkl} = \frac{1}{4} \Big\lbrack \, N(\cdots [[[j,k],i],l] \cdots; \ell) + N(\cdots [[[j,k],l],i] \cdots; \ell) \, \Big\rbrack \\
        & \qquad\qquad + \frac{1}{8} \Big\lbrack \, N(\cdots [[[i,j],k],l] \cdots; \ell) + N(\cdots [[[k,l],j],i] \cdots; \ell) \, \Big\rbrack \\
        & \qquad\qquad - \frac{1}{4}  N(\cdots [[i,j],[k,l]] \cdots; \ell) \,.
    \end{aligned}
    \label{eq:C7N}
\end{empheq}

We now impose monodromy invariance on the ansatz \eqref{eq:I7fay}. We have
\begin{equation}\label{eq:DiI7}
    \begin{aligned}
        0=&\; \cI_{7}|_{z_i \mapsto z_i + \tau}^{\ell\mapsto \ell - p_i} - \cI_7 \\
        =& \,
        -(2 \pi i)^3 \, M_{i}^{\mu\nu} \left( \ell_{\mu} \ell_{\nu} - \ell_{\mu} p_{i\nu} + \frac{1}{3}\, p_{i\mu} p_{i\nu} \right) \\
        & -2\, (\pi i)^2 \Bigg\lbrack \sum_{j \neq i} M_{ij}^{\mu} \Big( 2\, \ell_\mu\, g_{ij}^{(1)} - 2\pi i\, \ell_{\mu} - p_{i\mu}\, g_{ij}^{(1)} + \frac{4\pi i}{3} p_{i\mu} \Big) + \left( 2\, \ell_\mu - p_{i\mu} \right) \sum_{\substack{j,k\neq i \\ j<k}} M_{i;jk}^{\mu} g_{jk}^{(1)} \Bigg\rbrack \\
        &-2\pi i \Bigg\lbrack \sum_{\substack{j,k \neq i \\ j \neq k}} \hat{M}_{ijk} (g^{(1)}_{ij} - \pi i) \, g^{(1)}_{jk} 
        + \sum_{\substack{j,k \neq i \\ j < k}} \overline{M}_{ijk} \Big( g_{ij}^{(1)} g_{ik}^{(1)} - \pi i \, (g_{ij}^{(1)} + g_{ik}^{(1)}) - \frac{4 \pi^2}{3} \Big) \\
        & \qquad\qquad\quad
        + \sum_{\substack{j,k,l,\neq i \\ k<l,\, j\neq k,l}} M_{ij,kl} (g_{ij}^{(1)} - \pi i) \, g_{kl}^{(1)} + \sum_{j\neq i} \tilde{M}_{ij} \Big(\, g_{ij}^{(2)} - \pi i\, g_{ij}^{(1)} - \frac{2\pi^2}{3}\, \Big)  \Bigg\rbrack \\
        & -2\pi i \Bigg\lbrack \sum_{\substack{j,k\neq i \\ j<k}} \tilde{M}_{i;jk}\, g_{jk}^{(2)} 
        + \sum_{\substack{j,k,l \neq i \\ j<l,\, k\neq j,l}} M_{i;jkl}\, g_{jk}^{(1)} g_{lk}^{(1)} 
        + \sum_{\substack{j,k,l,m \text{\,dist\,} \neq i\\ j<k,l<m,j<l}} M_{i;jk,lm}\, g_{jk}^{(1)} g_{lm}^{(1)} \Bigg\rbrack \,.
    \end{aligned}
\end{equation}
We split the expression into: terms with $\ell$-dependence, terms with $g$-functions dependent on $z_i$, and terms with $g$-functions independent of $z_i$. The coefficients appearing in the first two lines, namely in the terms with $\ell$-dependence, are analogous to those in the 6-pt monodromy relation:
\begin{align}
\label{eq:M7iuv}
    M_{i}^{\mu\nu} &= 3\, p_{i\rho}\, C_7^{\mu\nu\rho} + \sum_{\substack{ i,j \\ i<j }}C^{\mu\nu}_{7,ij}  \,, \\
    M^{\mu}_{ij} &=  2\, p_{i\nu}\, C^{\mu\nu}_{7,ij} + \frac{1}{3} \sum_{k\neq i,j} \left( C^{\mu}_{7,j[ik]} + C^{\mu}_{7,k[ij]} \right) + \tilde{C}'^{\mu}_{7,ij}  \,, \\
    M^{\mu}_{i;jk} &=  2\, p_{i\nu}\, C^{\mu\nu}_{7,jk} + \sum_{l\neq i,j,k} C^{\mu}_{7,il,jk} - C^{\mu}_{7,i[jk]} \,.
    \label{eq:M7ijku}
\end{align}
Imposing monodromy invariance means that these coefficients must vanish. The remaining coefficients, appearing in the last three lines of \eqref{eq:DiI7}, are
\begin{align}
    \hat{M}_{ijk} &= -\frac{1}{3} \left( C^{\mu}_{7,i[jk]} + C^{\mu}_{7,k[ji]} \right) p_{i\mu} + \tilde{C}'_{7,jki} -\sum_{l \neq i,j,k} C'_{7,lijk} \,,\\
    \overline{M}_{ijk} &= -\frac{1}{3} \left( C^{\mu}_{7,j[ki]} + C^{\mu}_{7,k[ji]} \right) p_{i\mu} + \tilde{C}'_{7,ijk} + \tilde{C}'_{7,ikj} \\
    M_{ij,kl} &= C_{7,ij,kl}^{\mu} p_{i\mu} + C'_{7,jilk} - C'_{7,jikl} + \tilde{C}'_{7,kl,ij} + \frac{1}{3} \sum_{m \neq i,j,k,l} \left( C_{7,m[ij],kl} + C_{7,j[im],kl} \right) \,,\\
    \tilde{M}_{i;jk} &= \tilde{C}'^{\mu}_{7,jk}\, p_{i\mu} - \tilde{C}'_{7,jik} - \tilde{C}'_{7,kij} + \sum_{l\neq i} \tilde{C}'_{7,il,jk} \,, \\
    M_{i;jkl} &= \frac{1}{3} \left( C^{\mu}_{7,j[kl]} + C^{\mu}_{7,l[kj]} \right) p_{i\mu} -C'_{7,ijkl} - C'_{7,ilkj} + \frac{1}{3} \sum_{m\neq i,j,k,l} \left( C_{7, j[kl],im} + C_{7, l[kj],im} \right) \,, \\
    \tilde{M}_{ij} &= \tilde{C}'_{7,ij} + \tilde{C}'^{\mu}_{7,ij}\, p_{i\mu} + \sum_{k\neq i,j} \tilde{C}'_{7,ikj} \,.
\end{align}
Imposing monodromy invariance does not require that these coefficients must all vanish, because of the Fay relations. Instead, it requires that
\begin{equation}\label{eq:Lambda7}
    \begin{aligned}
        \cI_{7}|_{z_i \to z_i + \tau}^{\ell\to \ell - p_i} - \cI_7 =& \sum_{\substack{j,k \neq i \\ j<k}} \Lambda_{ijk} \left( g^{(2)}_{ik} + g^{(2)}_{ji} + g^{(2)}_{kj} +  g^{(1)}_{ij} g^{(1)}_{jk} + g^{(1)}_{jk} g^{(1)}_{ki} + g^{(1)}_{ki} g^{(1)}_{ij} \right) \\
        &+ \sum_{\substack{j,k,l \neq i \\ j<k<l}} \Lambda_{i;jkl} \left( g^{(2)}_{jl} + g^{(2)}_{kj} + g^{(2)}_{lk} +  g^{(1)}_{jk} g^{(1)}_{kl} + g^{(1)}_{kl} g^{(1)}_{lj} + g^{(1)}_{lj} g^{(1)}_{jk} \right)\,,
    \end{aligned}
\end{equation}
for some coefficients $\Lambda$. Matching this expression to the relevant terms in the monodromy relation \eqref{eq:DiI7}, the (sufficient and necessary) condition for $\Lambda$ to have solutions is
\begin{equation}\label{eq:7ptmonodromy}
    \begin{aligned}
        & M_{ij,kl} = 0 \,, \\
        & \hat{M}_{ijk} = \hat{M}_{ikj} = - \overline{M}_{ijk} \,, \qquad \tilde{M}_{ij} = \sum_{k \neq i,j} \hat{M}_{ijk} \,, \\
        & M_{i;jkl} = M_{i;klj} = M_{i;ljk} \,, \qquad \tilde{M}_{i;jk} = - \frac{1}{2} \left( \hat{M}_{ijk} + \hat{M}_{ikj} \right) - \sum_{l \neq i,j,k} M_{i;klj} \,.
    \end{aligned}
\end{equation}
Similarly to the 6-point case, it is useful to decompose these equations, together with the vanishing of the coefficients \eqref{eq:M7iuv}-\eqref{eq:M7ijku}, into two distinct (anti)symmetric sectors of monodromy constraints:
\begin{align}
    \tilde C \text{\;sector}\quad
    & \left\{
    \begin{aligned}
        & M_{(ij)}^\mu = 0 \,, \\
        & M_{(ij),kl} = 0 \,, \\
        & \hat{M}_{i[jk]} = 0 \,, \quad \hat{M}_{(ijk)} = - \overline{M}_{(ijk)} \,, \\
        & \tilde{M}_{[ij]} = \sum_{k\neq i,j} \hat{M}_{[ij]k} \,,
    \end{aligned}
    \right. \\ 
    \nonumber \\
    \text{BCJ sector}\quad
    & \left\{
    \begin{aligned}
        & M_{i}^{\mu\nu}=0\,, \quad M_{[ij]}^\mu = 0 \,, \quad M^{\mu}_{i;jk}=0\,, \\
        & M_{[ij],kl} = 0 \,, \\
        & \hat{M}_{i(jk)} + \overline{M}_{i(jk)} = \hat{M}_{j(ki)} + \overline{M}_{j(ki)} = \hat{M}_{k(ij)} + \overline{M}_{k(ij)}  \,, \\
        & M_{i;jkl} = M_{i;klj} = M_{i;ljk} \,.
    \end{aligned}
    \label{eq:7ptBCJsector}
    \right.
\end{align}
The top line in each sector is analogous to the 6-point case. Note also that some conditions in \eqref{eq:7ptmonodromy} are not included in either the $\tilde C$ sector or the BCJ sector, because they follow from the other conditions. In the case of \,$\tilde{M}_{(ij)} = \sum_{k \neq i,j} \hat{M}_{(ij)k}$\,, it can be rewritten as
\begin{equation}
    \tilde{M}_{(ij)} - \sum_{k \neq i,j} \hat{M}_{(ij)k} = M^{\mu}_{ij}\, p_{j\mu} + M^{\mu}_{ji}\, p_{i\mu} \,, 
\end{equation}
so it follows from the vanishing of $M^{\mu}_{ij}$. Likewise, the last condition in \eqref{eq:7ptmonodromy}, namely\, $ \tilde{M}_{i;jk} = - \frac{1}{2} ( \hat{M}_{ijk} + \hat{M}_{ikj} ) - \sum_{l \neq i,j,k} M_{i;klj} $\,, is not independent of the others. 

Now, the $\tilde C$ sector determines the `tilded' coefficients of our ansatz, which are associated with higher-weight $g$ functions. To be specific, the equations $M_{(ij)}^\mu = 0$, $M_{(ij),kl} =0$, $\hat{M}_{i[jk]} = 0$, $\hat{M}_{(ijk)} = -\overline{M}_{(ijk)}$, and $\tilde{M}_{[ij]} = \sum_{k\neq i,j} \hat{M}_{[ij]k}$ determine $\tilde{C}'^{\mu}_{7,ij}$, $\tilde{C}'_{7,ij,kl}$, $\tilde{C}'_{7,[ij]k}$, $\tilde{C}'_{7,(ijk)}$, and $\tilde{C}'_{7,ij}$, respectively; we are to combine $\tilde{C}'_{7,[ij]k}$ and $\tilde{C}'_{7,(ijk)}$ together into $\tilde{C}'_{ijk}$, according to \eqref{eq:newC7ijk}. The explicit expressions for the `tilded' coefficients in terms of those already given in \eqref{eq:C7N} are
\begin{equation}
    \boxed{
    \begin{aligned}
        \tilde{C}'^{\mu}_{7,ij} =& - (p_{i\nu}-p_{j\nu}) C^{\mu\nu}_{7,ij} - \frac{1}{6}\sum_{k\neq i,j}\, (C^{\mu}_{7,j[ik]}+C^{\mu}_{7,i[jk]}) \,, \\
        \tilde{C}'_{7,ij,kl} =& - \frac{1}{2} (p_{k\mu} - p_{l\mu}) C^\mu_{7,ij,kl} + \frac{1}{4} C'_{7,(kl)[ij]} - \frac{1}{6} \sum_{m \neq i,j,k,l} \left( C_{7,l[km],ij} + C_{7,k[lm],ij} \right) \,, \\
        \tilde{C}'_{7,ijk} =& \, \frac{1}{2} \sum_{l\neq i,j,k} C'_{7,lk[ij]} + \frac{1}{2} C^{\mu}_{7,k[ij]} p_{k\mu} + \frac{1}{18} \left( \sum_{l\neq i,j,k} C'_{7,lijk} - C^{\mu}_{7,k[ij]} p_{i\mu} + \text{perm} (i,j,k) \right) \,, \\
        \tilde{C}'_{7,ij} =& - \frac{1}{2} \tilde{C}'^{\mu}_{7,ij} ( p_{i\mu} - p_{j\mu} ) + \frac{1}{6} \sum_{k\neq i,j} \left( C^{\mu}_{7,i[kj]} + C^{\mu}_{7,j[ki]} \right)(p_j - p_i)_{\mu} \\
        & \; + \frac{1}{2} \sum_{\substack{k,l \neq i,j \\ k\neq l}} \left( C'_{7,ljki} - C'_{7,likj} \right) \,.
    \end{aligned}
    }
\label{eq:Ctilde7}
\end{equation}

The BCJ sector \eqref{eq:7ptBCJsector} then corresponds to the `BCJ monodromies', which are constraints on BCJ numerators, similarly to what we discussed at 5 and 6 points. In fact, the top line is precisely analogous to the 6-point case. The next constraint, $M_{[ij],kl} = 0$, corresponds to the case where the bi-particle $[i,j]$ goes around a loop which possesses a massive corner $[k,l]$. The constraint $\hat{M}_{i(jk)} + \overline{M}_{i(jk)} = \overline{M}_{j(ki)} + \hat{M}_{j(ki)}$ corresponds to the case where the tri-particle in the following goes around the loop with no other massive corners.
\begin{equation}
    \begin{tikzpicture}[scale=0.3,rotate=90]
        \draw[thick] (0,-2) -- (0,0) -- (-2,2) node[left]{$i$};
        \draw[thick] (0,0) -- (2,2) node[left]{$k$};
        \draw[thick] (-1,1) -- (0,2) node[left]{$j$};

        \draw[thick] (0,-4) circle (2);

        \draw[thick, fill=black] (-2.5,-3.4) circle (2pt); 
        \draw[thick, fill=black] (-2.5,-4) circle (2pt); 
        \draw[thick, fill=black] (-2.5,-4.6) circle (2pt); 

        \draw[thick, fill=black] (2.5,-3.4) circle (2pt); 
        \draw[thick, fill=black] (2.5,-4) circle (2pt); 
        \draw[thick, fill=black] (2.5,-4.6) circle (2pt);
    \end{tikzpicture}
    \nonumber
\end{equation}
The constraint $M_{i;jkl} = M_{i;ljk} = M_{i;klj}$ corresponds to the case where particle $i$ goes around a loop which possesses a massive corner including three particles $(j,k,l)$. For example, the equation $M_{i;jkl} = M_{i;ljk}$ corresponds to the loop containing a massive corner with the following tree.
\begin{equation}
    \begin{tikzpicture}[scale=0.3,rotate=-90]
        \draw[thick] (0,-2) -- (0,0) -- (-2,2) node[right]{$j$};
        \draw[thick] (0,0) -- (2,2) node[right]{$l$};
        \draw[thick] (-1,1) -- (0,2) node[right]{$k$};
        \draw[thick] (0,-6) -- (0,-8) node[left]{$i$};

        \draw[thick] (0,-4) circle (2);

        \draw[thick, fill=black] (-2.5,-3.4) circle (2pt); 
        \draw[thick, fill=black] (-2.5,-4) circle (2pt); 
        \draw[thick, fill=black] (-2.5,-4.6) circle (2pt); 

        \draw[thick, fill=black] (2.5,-3.4) circle (2pt); 
        \draw[thick, fill=black] (2.5,-4) circle (2pt); 
        \draw[thick, fill=black] (2.5,-4.6) circle (2pt);
    \end{tikzpicture}
    \nonumber
\end{equation}

To summarise this section, we have fully fixed the coefficients of our ansatz \eqref{eq:I7fay} in terms of BCJ numerators using the map (\ref{eq:C7N}, \ref{eq:Ctilde7}).

\section{Lessons for the $n$-point chiral integrand}
\label{sec:npt}

We have established a map between BCJ numerators and the superstring chiral integrand up to 7 points. In this section, we consider aspects of its extension to higher multiplicity.

\subsection{Leading-$\ell$ parts}

While the chiral integrands will be more complicated at higher points, we can already see some $n$-point structure arising from the previous examples.

Consider the $\ell$-dependent part of the 6-point chiral integrand \eqref{eq:I6fay}: it is structurally the same as the 5-point chiral integrand \eqref{eq:I5ansatz} with an additional factor $2\pi i \ell_\mu$, so at 6 points the associated coefficients acquire one more spacetime index. Then, the top line of the coefficient/numerator map \eqref{eq:C6N6} at 6 points is analogous to \eqref{eq:C5N} at 5 points.

Consider now the $\ell$-dependent part of the 7-point chiral integrand \eqref{eq:I7fay}: it is structurally the same as the 6-point chiral integrand \eqref{eq:I6fay} with an additional factor $2\pi i \ell_\mu$, so at 7 points the associated coefficients acquire one more spacetime index. Then, the top two lines of the coefficient/numerator map \eqref{eq:C7N} at 7 points are analogous to \eqref{eq:C6N6} at 6 points, while the top line of \eqref{eq:Ctilde7} at 7 points is analogous to \eqref{eq:C6N6tilde} at 6 points.

This is merely a consequence of the fixed weight of our ansatze. As we proceed up to $m$-points for some fixed $m$, we have already revealed how to fix the coefficients of the $\ell^{p\geq n-m}$ pieces of the chiral integrand at any $n>m$ points, in terms of the $n$-point BCJ numerators. Our paper sets the current status at $m=7$. To be fully explicit, the part of $\cI_n$ containing at least $n-7$ powers of the loop momentum is, using $\hat\ell_\mu=2\pi i\,\ell_\mu$,
\begin{empheq}[]{align}
    \cI_n|_{\ell^{p\geq n-7}}= \, & \hat\ell_{\lambda_1}\hat\ell_{\lambda_2}\cdots \hat\ell_{\lambda_{n-7}} \nonumber \\
    &
    \Bigg( C_n^{\lambda(n-7)\,\mu \nu \rho} \hat\ell_\mu \hat\ell_\nu \hat\ell_\rho
    +  \sum_{\substack{i, j \\ i<j}} C_{n, i j}^{\lambda(n-7)\,\mu \nu}\, \hat\ell_\mu \hat\ell_\nu\, g_{i j}^{(1)} 
    + \sum_{\substack{i, j, k, l \text { dist } \\ i<j, k<l, i<k}} \!C_{n, i j, k l}^{\lambda(n-7)\,\mu}\, \hat\ell_\mu \,g_{i j}^{(1)} g_{k l}^{(1)} \nonumber \\
    & +\frac{1}{3} \sum_{\substack{i, j, k \text { dist } \\
    i<k}}\left(C_{n, i[jk]}^{\lambda(n-7)\,\mu} + C_{n, k[j i]}^{\lambda(n-7)\, \mu}\right) \hat\ell_\mu \, g_{i j}^{(1)} g_{k j}^{(1)}
    + \sum_{\substack{i, j \\ i<j}} \tilde{C}_{n, i j}^{\prime \lambda(n-7)\,\mu}\, \hat\ell_\mu \,g_{i j}^{(2)} \nonumber \\
    & +\sum_{\substack{i, j, k, l, m, n \text { dist } \\ i<j, k<l, m<n, i<k<m}} C_{n, i j, k l, m n}^{\lambda(n-7)}\, g_{i j}^{(1)} g_{k l}^{(1)} g_{m n}^{(1)}
    \nonumber \\
    &
    - \frac{1}{3}\sum_{\substack{i, j, k, l, m \text { dist } \nonumber \\ j<k, l<m}} \left( C_{n, [i, j] k, l m}^{\lambda(n-7)} + C_{n, [i, k] j, l m}^{\lambda(n-7)} \right)\, g_{i j}^{(1)} g_{i k}^{(1)} g_{l m}^{(1)} \\
    & +\sum_{\substack{i, j, k, l \text { dist } \\ i<l}} C_{n, i j k l}^{\prime \lambda(n-7)}\, g_{i j}^{(1)} g_{j k}^{(1)} g_{k l}^{(1)}
    + \sum_{\substack{i, j, k, l \text { dist } \nonumber \\ i <j, k<l}} \tilde{C}_{n, i j, k l}^{\prime\lambda(n-7)}\, g_{i j}^{(1)} g_{k l}^{(2)} \\
    & +\sum_{i, j, k \text { dist }} \tilde{C}_{n, i j k}^{\prime\lambda(n-7)}\, g_{i j}^{(1)} g_{k i}^{(2)}
    + \sum_{\substack{i, j \\ i<j}} \tilde{C}_{n, i j}^{\prime\lambda(n-7)} \,g_{i j}^{(3)} \Bigg)\,,
    \label{eq:Infay}
\end{empheq}
where $\lambda(n-7)$ is a shorthand notation for spacetimes indices, such that $C^{\lambda(n-7)\,\cdots}=C^{\lambda_1\lambda_2\cdots \lambda_{n-7}\,\cdots}$. For instance, the first term in this expression is $\hat\ell_{\lambda_1}\cdots\hat\ell_{\lambda_{n-4}} C_n^{\lambda_1\cdots\lambda_{n-4}}$. Notice the similarity of the expression above to the 7-point chiral integrand \eqref{eq:I7fay}.
The fixing of the various coefficients here in terms of $n$-point BCJ numerators follows exactly the same pattern that we saw at 7 points. In particular, the map (\ref{eq:C7N}, \ref{eq:Ctilde7}) applies here with only minor alterations. The coefficients on the left-hand side in the map \eqref{eq:C7N} should be appropriately renamed, while on the right-hand side, the piece of the numerators with the appropriate power of $\ell$ should be extracted. For instance, the $n$-point analogue of the 7-point identification 
\[
C_{7,ij,kl}^{\mu}  =  N(\cdots[i,j]\cdots[k,l]\cdots;\ell)|_{\ell_\mu}\,,
\]
is simply
\[
C_{n, i j, k l}^{\lambda_1\lambda_2\cdots \lambda_{n-6}}  =  N(\cdots[i,j]\cdots[k,l]\cdots;\ell)|_{\ell_{\lambda_1}\ell_{\lambda_2}\cdots\ell_{\lambda_{n-6}}}\,.
\]
As for the \eqref{eq:Ctilde7} part of the 7-point map, corresponding to the `tilded' coefficients, the analogous part of the $n$-point map is similar, though various numerical factors are $n$-dependent:
\begin{equation}
        \begin{aligned}
            \tilde{C}'^{\mu_1\cdots \mu_{n-6}}_{n,ij} =& - \frac{n-5}{2} (p_{i\nu}-p_{j\nu}) C^{\mu_1\cdots \mu_{n-6}\nu}_{n,ij} - \frac{1}{6}\sum_{k\neq i,j}\, (C^{\mu_1 \cdots \mu_{n-6}}_{n,j[ik]}+C^{\mu_1 \cdots \mu_{n-6}}_{n,i[jk]}) \,, \\
            \tilde{C}'^{\mu_1 \cdots \mu_{n-7}}_{n,ij,kl} =& - \frac{n-6}{2} (p_{k\mu} - p_{l\nu}) C^{\mu_1 \cdots \mu_{n-7}\nu}_{n,ij,kl} + \frac{1}{4} C'^{\mu_1 \cdots \mu_{n-7}}_{n,(kl)[ij]} - \frac{1}{6} \sum_{m \neq i,j,k,l} \left( C^{\mu_1 \cdots \mu_{n-7}}_{n,l[km],ij} + C^{\mu_1 \cdots \mu_{n-7}}_{n,k[lm],ij} \right) \,, \\
            \tilde{C}'^{\mu_1 \cdots \mu_{n-7}}_{n,ijk} =& \, \frac{1}{2} \sum_{l\neq i,j,k} C'^{\mu_1 \cdots \mu_{n-7}}_{n,lk[ij]} + \frac{n-7}{2} C^{\mu_1 \cdots \mu_{n-7}\nu}_{n,k[ij]} p_{k\nu} \\
            & + \frac{1}{18} \left( \sum_{l\neq i,j,k} C'^{\mu_1 \cdots \mu_{n-7}}_{n,lijk} - (n-7)C^{\mu_1 \cdots \mu_{n-7}\nu}_{n,k[ij]} p_{i\nu} + \text{perm} (i,j,k) \right) \,, \\
            \tilde{C}'^{\mu_1 \cdots \mu_{n-7}}_{n,ij} =& - \frac{n-7}{2} \tilde{C}'^{\mu_1 \cdots \mu_{n-7}\nu}_{n,ij} ( p_{i\nu} - p_{j\nu} ) + \frac{n-7}{6} \sum_{k\neq i,j} \left( C^{\mu_1 \cdots \mu_{n-7}\nu}_{n,i[kj]} + C^{\mu_1 \cdots \mu_{n-7}\nu}_{n,j[ki]} \right)(p_j - p_i)_{\nu} \\
            & \; + \frac{1}{2} \sum_{\substack{k,l \neq i,j \\ k\neq l}} \left( C'^{\mu_1 \cdots \mu_{n-7}}_{n,ljki} - C'^{\mu_1 \cdots \mu_{n-7}}_{n,likj} \right) \,.
        \end{aligned}
\end{equation}

The question that we leave open is whether the coefficients of the $\ell^{p< n-7}$ pieces of the $n$-point chiral integrand can also be fixed in the same manner, using only the degeneration limit and the monodromy constraints. There is a competition between weaker information about the degeneration limit due to \eqref{eq:ghigherlim}, on the one hand, and stronger constraints coming from monodromy, on the other hand.

\subsection{Cusp forms at high multiplicity}

So far, we did not discuss the appearance of the monodromy-blind objects $G_{2K}(\tau)$ in chiral integrands. Recalling table \eqref{eq:tablew}, these objects will be part of the ansatze starting at 8 points ($K\geq 2$), and they obey $G_{2K}(\tau)\to 2 \,\zeta(2K)$ as $q\to 0$. Since their degeneration limit is non-vanishing, the coefficients associated to these objects will contribute in the field-theory limit. For instance, at 8 points the coefficient of $G_4(\tau)$ will contribute to the permutation-symmetric $\ell$-independent part of the BCJ numerators. There is in our current understanding no obvious obstruction to fixing all coefficients for $n<16$ points (weight$~<12$), though this is under investigation. There is, however, a clear obstruction arising at 16 points and beyond, due to the possible appearance of cusp forms. For example, at 16 points we have modular weight 12, and all modular forms with this weight are a linear combination of $(G_4)^3$ and $(G_6)^2$. Hence, one way of expressing that contribution to the chiral integrand $\cI_{16}$ is
\[
\label{eq:G4G6}
C_{16,{G_4}}\, G_4(\tau)^3 + C_{16,{G_6}}\, G_6(\tau)^2\,,
\]
where $C_{16,{G_4}}$ and $C_{16,{G_6}}$ are functions of the external scattering data $\{p_i,\epsilon_i\}$. However, only a linear combination of these two coefficients can be fixed by the degeneration limit, due to the existence of the weight-12 cusp form
\[
\label{eq:Gcusp}
G_\text{cusp} = (2\pi)^{12}\left[ \left( \frac{G_4(\tau)}{2\, \zeta(4)} \right)^3 - \left( \frac{G_6(\tau)}{2\, \zeta(6)} \right)^2 \right] \,.
\]
By cusp form, we mean that this modular form vanishes as $q\to0$, so we miss some information about the coefficients in \eqref{eq:G4G6} in the degeneration limit. If the notion that the superstring amplitude is just field theory dressed with $\alpha'$ is correct, then the missing information must somehow be trivial. If this is actually the case, it is unclear to us at present how such triviality manifests itself. To illustrate how it is conceivable, consider the following expression that is equivalent to \eqref{eq:G4G6} up to redefinition of the $C$ coefficients:
\[
\label{eq:G12Gcusp}
C_{16,{G_{12}}}\, G_{12}(\tau) + C_{16,{G_\text{cusp}}}\, G_\text{cusp}(\tau)\,.
\]
If it turns out that only $G_{12}$ appears in the superstring integrand, then this would mean that $C_{16,{G_\text{cusp}}}=0$. We cannot reach this conclusion merely from the field-theory limit, which is blind to $C_{16,{G_\text{cusp}}}$. So, in order to obtain all the information about the superstring from the field-theory limit --- if that is indeed possible --- we would need a refinement of the ansatz presented in section~\ref{sec:structure}. An obvious option is that the ansatz would allow only for the appearance of the holomorphic Eisenstein series, $G_{2K}(\tau)$, as in the example above. At each modular weight $2K$ (modular forms must have even weight), any modular form is a linear combination of $G_{2K}(\tau)$ and cusp forms, so in this scenario the coefficients of all cusp forms in this basis would vanish. This is just an illustration of what could happen, and we have no evidence that this is the case for superstring amplitudes. Recent work  \cite{Dorigoni:2021ngn} on modular graph forms may also be relevant to this discussion: it was seen there that sets of cusp forms appear in a special way in intermediate steps of the construction such that their overall contribution to a modular graph form vanishes.

To conclude, we noted the obstruction to fixing all the coefficients of the ansatz presented in section~\ref{sec:structure} if we allow for all possible module forms, i.e.~any polynomial of $G_4$ and $G_6$ with appropriate weight. We noted also that it is still conceivable that a natural ansatz exists where all coefficients can be fixed, e.g.~if at a given weight $2K$ only $G_{2K}$ is allowed in the ansatz. Whether the superstring chiral integrand obeys such a restricted ansatz is an open question, which is a translation in our framework of the question of whether the superstring amplitude can be fully constructed using information from the field-theory limit.

\section{Some explicit checks}
\label{sec:checks}

In this section, we will illustrate the results with explicit formulas for one-loop BCJ numerators taken from \cite{Edison:2022jln}, focusing on external massless states of the NS-NS (NS) type for the closed (open) superstring. The numerators will depend on the external momenta $p_i^\mu$, the associated polarisations $\epsilon^\mu_i$, and the loop momentum $\ell_\mu$. In particular, we will perform checks of gauge invariance.

Following \cite{Edison:2022jln}, which builds on previous work using the pure-spinor formalism \cite{Mafra:2014oia,Mafra:2014gja,Mafra:2015vca}, it is useful to define multiparticle momenta, polarisations and field strengths. For the momenta, we have
\[
p_{12\ldots m}=p_1 +p_2+\ldots +p_m\,, \qquad 
\ell_{12\ldots m} = \ell + p_1 + p_2+\ldots +p_m\,.
\]
For the polarisations and field strengths, we start with the single-particle case,
\[
f_i^{\mu\nu} = p_i^\mu \epsilon_i^\nu - \epsilon_i^\mu p_i^\nu\,,
\]
and proceed to the multi-particle cases: up to the order we will use here explicitly,
\begin{align}
    \epsilon_{12}^{\mu}&=(\epsilon_1\cdot p_2)\epsilon_2^{\mu}-(\epsilon_2\cdot p_1)\epsilon_1^{\mu}+\frac{1}{2}(\epsilon_1\cdot\epsilon_2)(p_1-p_2)^{\mu}\,,\nonumber\\
    f_{12}^{\mu\nu} &= p_{12}^{\mu}\epsilon_{12}^{\nu}- p_{12}^{\nu}\epsilon_{12}^{\mu}-(p_{1}\cdot p_2)(\epsilon_{1}^{\mu}\epsilon_2^{\nu}-\epsilon_{1}^{\nu}\epsilon_2^{\mu})\,,
\end{align}
and
\begin{align}
\epsilon_{123}^{\mu}&=\frac{1}{2}\Big[(\epsilon_{12}\cdot p_3)\epsilon_3^{\mu}-(\epsilon_3\cdot p_{12})\epsilon_{12}^{\mu}+\epsilon_{12,\nu}f_3^{\nu\mu}-\epsilon_{3,\nu}f_{12}^{\nu\mu}\Big] -p_{123}^{\mu}h_{123}\,,
\notag \\
f_{123}^{\mu\nu}&=p_{123}^{\mu}\epsilon_{123}^{\nu}-(p_{12}\cdot p_3)\epsilon_{12}^{\mu}\epsilon_3^{\nu}-(p_1\cdot p_2)(\epsilon_{1}^{\mu}\epsilon_{23}^{\nu}-\epsilon_2^{\mu}\epsilon_{13}^{\nu})-(\mu\leftrightarrow \nu)\,,
\end{align}
where\, $h_{123}=\frac{1}{12}\epsilon_{1\mu}f_2^{\mu\nu}\epsilon_{3\nu}+ \text{cyc}(1,2,3)$\,. We will also consider trace-type contractions,  such as $\tr(f_if_j)=f_{i\mu}{}^\nu f_{j\nu}{}^\mu$.

\subsection{4 points}
\label{sec:4ptexplicit}

The manifestly gauge-invariant expression
\[
t_8(f_1,f_2,f_3,f_4) = 
\tr(f_1 f_2 f_3 f_4)
- \frac{1}{4} \tr(f_1 f_2 ) \tr(f_3 f_4) + \text{cyc}(2,3,4)
\]
is permutation-symmetric in the particle labels. The 4-point BCJ numerators are
\[
N(1,2,3,4;\ell)=t_8(f_1,f_2,f_3,f_4) \,,
\]
corresponding to the long-known result \cite{Green:1981yb,Schwarz:1982jn,Green:1982sw}
\[
\cI_4= t_8(f_1,f_2,f_3,f_4) \,.
\]

\subsection{5 points}

The 5-point master BCJ numerators can be written as \cite{Edison:2022jln}
\begin{align}
\label{eq:5ptEdison}
    N(12345;\ell)&=\ell_{\mu} t_8^{\mu}(1,2,3,4,5)-\frac{1}{2}\Big[t_8(f_{12},f_3,f_4,f_5)+(1,2|1,2,3,4,5)\Big]\nonumber\\
    &\quad  + \frac{1}{16}\, \varepsilon_{10}(\ell_1,\epsilon_1,f_2,f_3,f_4,f_5)\,,
\end{align}
where we denote
\[
t_8^{\mu}(1,2,3,4,5)&=\epsilon_1^{\mu} t_8(f_2,f_3,f_4,f_5)+\text{cyclic}(1,2,3,4,5)\,.
\]
The latter object is permutation-symmetric.
The notation $(1,2|1,2,3,4,5)$ represents a sum over pairs $(i,j)$ obeying the ordering $(1,2,3,4,5)$, so that there are 10 terms inside the brackets $[\cdots]$ in \eqref{eq:5ptEdison}. We note that, in contrast to the 4-point case, we now have a parity-odd contribution, which is associated to the odd spin structure in the RNS worldsheet correlator. This contribution is also permutation-symmetric due to momentum conservation.

Using the map \eqref{eq:C5N}, we can read off
\begin{align}
C_5^\mu & = N(12345;\ell)|_{\ell_\mu} = t_8^{\mu}(1,2,3,4,5) + \frac{1}{16} \,\varepsilon_{10}^\mu(\cdot,\epsilon_1,f_2,f_3,f_4,f_5)\,, \nonumber \\
C_{5,12} & = - N([1,2]345;\ell) = t_8(f_{12},f_3,f_4,f_5)\,,
\end{align}
determining the explicit chiral integrand $\cI_5$ in \eqref{eq:I5ansatz}.

A basic check is gauge invariance. Setting $\epsilon_1=p_1$, and recalling the chiral Koba-Nielsen factor \eqref{eq:KN}, we obtain
\[
\cI_5|_{\epsilon_1=p_1}\, \text{KN}_5 = \Bigg(2\pi i\, \ell\cdot p_1 + \sum_{j>1} p_1\cdot p_j \, g^{(1)}_{1j}\Bigg)
\, t_8(f_2,f_3,f_4,f_5)\, \text{KN}_5 = 
\frac{2}{\alpha'} \,t_8(f_2,f_3,f_4,f_5)\, \frac{\partial}{\partial z_1} \text{KN}_5\,.
\]
Hence, the pure-gauge moduli-space integrand is a total derivative, as expected.

\subsection{6 points}

A complete expression for the 6-point master BCJ numerator is provided in \cite{Edison:2022jln}. The parity-even part is 
\begin{align}
    N^\text{even}(123456;\ell)&= \Big[ \epsilon_1\cdot\ell_{1}\,\epsilon_2\cdot\ell_{2}
   \,t_8(f_3 f_4 f_5 f_6)+(1,2|1,2,3,4,5,6)\Big] \nonumber \\
    &\quad - \Big[\epsilon_1\cdot\ell_1 \, t_8'(f_2 f_3 f_4 f_5 f_6)+\text{cyc}(1,2,3,4,5,6)\Big]+t_8''(f_1 f_2 f_3 f_4 f_5f_6)\nonumber\\
    &\quad  +\frac{1}{40}\Big[\epsilon_1\cdot\epsilon_2 \, (3\ell^2_{2}-10\ell^2_{1}+3\ell^2_{6})\,t_8(f_3f_4f_5f_6)\nonumber\\
    &\qquad\quad\quad+\epsilon_1\cdot\epsilon_3 \, (\ell^2_{3}-3\ell^2_{2}-3\ell^2_{1}+\ell^2_{6})\,t_8(f_2f_4f_5f_6)\nonumber\\
    &\qquad\quad\quad-\epsilon_{1}\cdot\epsilon_4 \, (\ell^2_{1}+\ell^2_{6})\,t_8(f_2f_3f_5f_6)
     +\text{cyc}(1,2,3,4,5,6)  \Big]\,,
\end{align}
where
\begin{align}
t_8'(f_1 f_2 f_3 f_4 f_5)&= \frac{1}{2}\, t_8(f_1,[f_2,f_3],f_4,f_5)+(2,3|2,3,4,5)\,,  \\
t_8''(f_1 f_2 f_3 f_4 f_5 f_6)&= \frac{1}{24}\Big[ \tr(f_1f_2)\,t_8(f_3,f_4,f_5,f_6)+(1,2|1,2,3,4,5,6)\Big]
\nonumber \\
&\quad+
\frac{1}{6}\Big[t_8(f_1,
[[f_2,f_3],f_4],f_5,f_6)+t_8(f_1,
[[f_4,f_3],f_2],f_5,f_6)+(2,3,4|2,3,4,5,6)\Big]
\nonumber \\
&\quad+ \frac{1}{4}\Big[t_8(f_1,
[f_2,f_3],[f_4,f_5],f_6)
    +t_8(f_1,
    [f_2,f_4],[f_3,f_5],f_6)\nonumber \\
&\quad +t_8(f_1,
    [f_2,f_5],[f_3,f_4],f_6)+(2,3,4,5|2,3,4,5,6)\Big]\,.
\end{align}
The parity-even part of the chiral correlator, $\cI_6^\text{even}$, follows from the map (\ref{eq:C6N6}, \ref{eq:C6N6tilde}).
Setting $\epsilon_1=p_1$ to check gauge invariance, we find\footnote{To determine $\partial_\tau  \text{KN}_6$, one should note the following identity:\, $4\pi i \,\partial_\tau \theta_1(z,\tau) = \partial_z^2 \theta_1(z,\tau) $\,.}
\begin{align}
    \cI_6|_{\epsilon_1=p_1}\, \text{KN}_6 & = \,
    \chi \Bigg((2\pi i)^2\, \ell^2 + 2\, \sum_{\substack{i,j\\ i<j}}\, p_i\cdot p_j\, g^{(2)}_{ij} \Bigg)\text{KN}_6 \nonumber \\
    &\qquad + \sum_{i=1}^6
    \Bigg(2\pi i\, \ell_\mu\, \chi^\mu_i + \sum_{\substack{j,k\neq i\\  j<k}}  \chi_{i,jk}\,g^{(1)}_{jk}\Bigg)
    \Bigg(2\pi i\, \ell\cdot p_i + \sum_{l\neq i} p_i\cdot p_l \, g^{(1)}_{il}\Bigg)
    \text{KN}_6 =\nonumber \\
    & = \chi\, \frac{8\pi i}{\alpha'} \, \frac{\partial}{\partial \tau}\,  \text{KN}_6 \nonumber \\
    &\qquad + \frac{2}{\alpha'} \sum_{i=1}^6 \Bigg(2\pi i\, \ell_\mu\, \chi^\mu_i + \sum_{\substack{j,k\neq i\\  j<k}}  \chi_{i,jk}\,g^{(1)}_{jk}\Bigg)\frac{\partial}{\partial z_i}\,  \text{KN}_6 \,,
    \label{eq:I6k1}
\end{align}
which is, as expected, a total derivative in moduli space; note that the terms inside the brackets in the last line are independent of $z_i$. To be clear, the first and second lines after the first equality in \eqref{eq:I6k1} correspond to the first and second lines after the second equality, respectively. The complete expression is invariant under monodromy \eqref{eq:monodromy}, because $\cI_6$ is invariant, for generic polarisations, and so is $\text{KN}_6$. However, the two lines in the expression above are not separately monodromy-invariant.\footnote{Notice that, while $\text{KN}_6$ is monodromy invariant, $\partial_\tau$ is not. Under $(\tau,z_i)\mapsto(\tau,z_1+\tau,z_{i>1})$, we have $\partial_\tau\mapsto \partial_\tau+\partial_{z_1}$.}
Explicitly, the coefficients in \eqref{eq:I6k1}, which depend only on the external kinematics, are
\[
\chi = -\frac1{10} \,\sum_{j>1} p_1\cdot \epsilon_j \, t_8(\text{not }1,j)\,,
\qquad
\chi^\mu_1 = \sum_{j>1} \epsilon_j^\mu \, t_8(\text{not }1,j)\,,
\qquad
\chi^\mu_{i>1} =0\,,
\]
and
\begin{align}
    \chi_{1,jk} & = \frac1{5} \, p_1\cdot\left[ \epsilon_j   \, t_8(\text{not }1,j) - \epsilon_k   \, t_8(\text{not }1,k)\right] +t_8(f_{jk}, \text{not }1,j,k)\,, \nonumber \\
    \chi_{i,1k} & = \frac1{10}\, p_1\cdot \left[ \epsilon_i\, t_8(\text{not }1,i) + 4\, \epsilon_k\, t_8(\text{not }1,k) + \sum_{l>1} \epsilon_l\, t_8(\text{not }1,l) \right]\,,
     \nonumber \\
    \chi_{i,jk}|_{i,j,k\neq 1} & = \frac1{10} \, p_1\cdot \left[ \epsilon_j\, t_8(\text{not }1,j) - \epsilon_k\, t_8(\text{not }1,k) \right]\,, \forall {i> 1}\,.
\end{align}
We denote, for instance, $\,t_8(\text{not }1,3)=t_8(f_2,f_4,f_5,f_6)\,$, and $\,t_8(f_{35}, \text{not }1,3,5)=t_8(f_{35},f_2,f_4,f_6)\,$. Recall also that $t_8(f_i,f_j,f_k,f_l)$ is invariant under permutations of $\{i,j,k,l\}$, and we also have that $t_8(f_{ij},f_k,f_l,f_m)$ is antisymmetric in $\{i,j\}$ and invariant under permutations of $\{k,l,m\}$.

For completeness, we write here the parity-odd part of the BCJ numerators \cite{Edison:2022jln}:
\begin{align}
    N^\text{odd}(123456;\ell)&=\frac{1}{96}\Big[\epsilon_{2}\cdot\ell_2 \, \varepsilon_{10}(\ell_1,\epsilon_{1},f_3,f_4,f_5,f_6)\nonumber\\
    &\qquad\quad+\epsilon_{1}\cdot\ell_1 \, \varepsilon_{10}(\ell_2,\epsilon_{2},f_3,f_4,f_5,f_6)+(1,2|1,2,3,4,5,6)\Big]\nonumber\\
    &\quad+\frac{1}{384}\Big[\epsilon_{1}\cdot\ell_1 \, \varepsilon_{10}(f_2,f_3,f_4,f_5,f_6)+\textrm{cyc}(1,2,3,4,5,6)\Big]\nonumber\\
    &\quad -\frac{1}{192}\Big[\Big(\varepsilon_{10}(\ell_3,\epsilon_{3},[f_1,f_2],f_4,f_5,f_6)+\textrm{cyc}(1,2,3)\Big)+(1,2,3|1,2,3,4,5,6)\Big]\nonumber\\
    &\quad +\frac{1}{64}\Big[\ell^2_{1} \, \varepsilon_{10}(\epsilon_{1},\epsilon_{2},f_3,f_4,f_5,f_6)+\textrm{cyc}(1,2,3,4,5,6)\Big]\,.
\end{align}
The associated parity-odd contribution to the chiral correlator $\cI_6$ is fully determined by our map (\ref{eq:C6N6}, \ref{eq:C6N6tilde}). We will not analyse here the pure-gauge case, which is not as straightforward as for the parity-even part, due to the gauge anomaly.

\subsection{7 points}

 Our map fully determines $\cI_7$ from the 7-point master BCJ numerators. The parity-even part of these numerators was explicitly given in section 4.4 of \cite{Edison:2022jln}. After looking at those numerators, and considering that the checks of gauge invariance we performed for 5 and 6 points were mostly done by hand, we have decided not to perform such a check at 7 points at this stage.

\section{Conclusion}
\label{sec:conclusion}

We have found, up to 7 points, a map that allows for the construction of the superstring moduli-space integrand at one loop from the knowledge of the field-theory loop integrand in BCJ form, i.e.~satisfying the colour-kinematics duality. In fact, this map extends to the `leading-$\ell$' parts of the superstring moduli-space integrands at any multiplicity. The map relies on a combination of (i) the degeneration limit corresponding to field theory, and (ii) the monodromy constraints that are part of the chiral-splitting representation. It is remarkable that all the coefficients that cannot be fixed by the degeneration limit are rescued by monodromy invariance. The 7-point case that we presented in detail illustrates how non-trivial this is. While the expressions inevitably become longer at higher points, we have revealed an unexpected structural simplicity.

No obstacle is known in constructing BCJ numerators for maximal super-Yang-Mills and supergravity at one loop. In fact, our ansatze for the superstring moduli-space integrand imply that BCJ numerators exist. This is irrespective of whether or not all the coefficients of the ansatze are fixed by our method from the knowledge of BCJ numerators at higher multiplicity.

For clarity on future investigations at higher multiplicity, let us consider the different possibilities given our results.
\begin{itemize}
    \item Option A: the ansatze we presented in section~\ref{sec:structure} are complete. We have already established that all the coefficients can be fixed up to 7 points. We have also established, however, that not all coefficients can be fixed starting at 16 points, due to the appearance of cusp modular forms starting at modular weight 12. These cusp forms are monodromy invariant and vanish in the degeneration limit associated to field theory, so their coefficients in the ansatz cannot be fixed in the manner we described without further input. We have two options.
    \begin{itemize}
    \item Option A1: all coefficients of a refined ansatze can be fixed by a combination of degeneration limit and monodromy invariance. This would require a very stringent restriction on the modular forms. One possibility is that, for any modular weight $2K$, only the holomorphic Eisenstein series $G_{2K}(\tau)$ would arise, as illustrated in section~\ref{sec:npt}. There is no evidence at present that this is the case, although non-trivial cancellations of cusp forms occur in similar contexts \cite{Dorigoni:2021ngn}. Option A1 would mean that the superstring moduli-space integrand is just an $\alpha'$-dressing of field theory.
    
    \item Option A2: not all coefficients of a natural ansatz can be fixed solely by a combination of degeneration limit and monodromy invariance. This could be because the issue of cusp forms at high multiplicity is unfixable; but it could also be because the coefficients cannot be determined at some not-so-high multiplicity, before the occurrence of cusp forms, although at present we see no obvious obstruction. The latter question is easier to investigate in the near future, if we apply our approach at 8 points and beyond.
    \end{itemize}
    \item Option B: the ansatze we presented are not complete.
    Expressions for the 6-point moduli-space integrand that were previously obtained from the pure-spinor formalism do not obey our ansatz directly, due to presence of $\partial_{z_i}g^{(1)}_{ij}$, but it was shown in parallel work \cite{Balli:2024wje} that careful integration by parts in moduli space addresses this issue. We certainly expect that our 7-point ansatz is complete, given how tight the structure we found is. To clarify this, we need a cohomological understanding of the basis at genus 1 analogous to the current understanding at genus 0.
\end{itemize}

There are other questions that we leave open. It was clear how to use our ansatze in the cases we studied, but at higher points a better understanding of what terms can be excluded, and of how to apply the Fay relations efficiently would be useful. This is closely connected to the comments above in Option B. A different point is that it should be possible to demonstrate explicitly that the numerators $N$ appearing in the degeneration formula \eqref{eq:assumpq0} are indeed identified -- to any multiplicity -- with the BCJ numerators in the field-theory limit. One intriguing aspect that we mentioned at the end of section~\ref{sec:6pt} is that the BCJ numerators are more constrained than the naive field-theory expectation starting at 6 points. The string-theory counterpart of this constraint is the absence of $G_{2}(\tau)$ from the ansatz, due to the fact that, while the Eisenstein series converges in this case, it is not a modular form.

Yet another natural question is whether our story can lead to simplifications of the final amplitude, i.e.~after moduli-space integration. The $\alpha'$ expansion (see e.g.~\cite{Broedel:2018izr,Edison:2021ebi} for recent work) is an obvious direction. Moreover, one-loop 4-point amplitudes have recently been evaluated numerically \cite{Eberhardt:2023xck} and their unitarity properties revisited \cite{Eberhardt:2022zay}, after a new understanding of the integration contour. The structure we unveiled is helpful for higher-point amplitudes, and it explicitly ties the string-theory properties to the underlying field-theory ones.

We are clearly seeing a novel structure of interest to superstring amplitudes and their field-theory limit, and perhaps also of interest to the mathematical study of Riemann surfaces. The power of this approach had already been seen in the conjecture for the 3-loop 4-point amplitude \cite{Geyer:2021oox}, which -- to date -- no worldsheet CFT technique has been able to address fully. While in the present paper we focused on 1 loop, existing 2-loop results provide a substantial playing ground, beyond the 4-point case already dealt with in \cite{Geyer:2021oox}. In particular, the 2-loop 5-point superstring amplitude has been constructed in \cite{DHoker:2020prr,DHoker:2020tcq,DHoker:2021kks}. Moreover, there has been important progress in obtaining analogues of the torus $g$-functions we used, based on \cite{brown2013multipleellipticpolylogarithms},
at higher genus \cite{DHoker:2022xxg,DHoker:2023vax,DHoker:2023khh,DHoker:2024ozn}. Indeed, in the 3-loop conjecture of \cite{Geyer:2021oox}, analogous objects corresponding to RNS-type sums over spin structures at genus 3 were essential \cite{DHoker:2004fcs,DHoker:2004qhf,Cacciatori:2008ay,Belavin:1986tv}; see also \cite{Grushevsky:2008zm,Cacciatori:2008pj,SalvatiManni:2008qa,Morozov:2008wz,Grushevsky:2008zp,Matone:2010yv,Matone:2005vm}. On the field-theory side, well-established progress has reached 5 loops \cite{Bern:2017ucb,Bern:2018jmv}, with some work also at 6 loops \cite{Carrasco:2021otn}, and there are significant recent developments \cite{Bern:2024vqs,Edison:2023ulf}. It is tempting to anticipate that the relaxation of the BCJ structure needed at 5 loops \cite{Bern:2017ucb} is related to the non-projectability of supermoduli space \cite{Donagi:2013dua}, which may in turn be associated to a failure of global chiral splitting for the amplitudes. In any case, we believe that these questions could be addressed in the not-too-distant future, and that our approach may provide an alternative to -- or at least complement -- the worldsheet CFT machinery.

\subsection*{Acknowledgements}

We are especially grateful to Oliver Schlotterer for several discussions and for comments on the manuscript, and also thank Filippo Balli, Alex Edison and Oliver Schlotterer for sharing their draft with us in advance of publication. In addition, we thank Erik D'Hoker, Henrik Johansson, Rodolfo Russo and Congkao Wen for discussions. YG and RM are grateful to the organisers and participants of the workshop String Amplitudes at Finite $\alpha'$, at IAS Princeton in 2023. RM and LR are supported by the Royal Society, via a University Research Fellowship and an associated postdoctoral grant, respectively. This work is also supported by the UK's Science and Technology
Facilities Council (STFC) Consolidated Grants ST/T000686/1 and ST/X00063X/1 ``Amplitudes, Strings \& Duality".

\appendix

\section{Details on 6-point `BCJ monodromies'}
\label{app:6ptBCJmonodromies}

This appendix includes details on how the monodromy constraints for the coefficients of the 6-point superstring ansatz relate to the constraints among BCJ numerators, complementing the discussion a bit below equation \eqref{eq:C6N6tilde}. This is analogous to the simpler 5-point case presented in \eqref{eq:N5monodromy}.

The vanishing of the coefficients $M_i^\mu$ defined in \eqref{eq:Mmui6} corresponds to, say for $i=1$,
\begin{align}
& 2\,p_{1\nu}C_6^{\mu\nu} =\big( N(123456;\ell)- N(123456;\ell-p_1)\big)|_{\ell_\mu} = 
\nonumber \\
&\quad = \big( N(123456;\ell)- N(234561;\ell)\big)|_{\ell_\mu} =
\nonumber \\
&\quad =\big( N([1,2]3456;\ell) + N(2[1,3]456;\ell) +N(23[1,4]56;\ell) + N(234[1,5]6;\ell)+ N(2345[1,6];\ell)\big)|_{\ell_\mu} =
\nonumber \\
&\quad = - \sum_{j>1}\, C^\mu_{6,1j}\,,
\label{eq:Mimuapp}
\end{align}
where we employed in the first and last equality the map \eqref{eq:C6N6}. So we are taking particle 1 around the hexagon.

The vanishing of the coefficients $M_{[ij]}$ defined in \eqref{eq:M6ijanti} corresponds to, say for $\{i,j\}=\{1,2\}$,
\begin{align}
& (p_{1\mu}+p_{2\mu})C_{6,ij}^{\mu} = - N([1,2]3456;\ell)+ N([1,2]3456;\ell-p_1-p_2) = \nonumber \\
&\quad = - N([1,2]3456;\ell)+ N(3456[1,2];\ell) =
\nonumber \\
&\quad = -N([[1,2],3]456;\ell) - N(3[[1,2],4]56;\ell) - N(34[[1,2],5]6;\ell) - N(345[[1,2],6];\ell) =
\nonumber \\
&\quad = - \sum_{k>2}\, C_{6,k[1,2]}\,,
\end{align}
where we employed in the first and last equality the map \eqref{eq:C6N6}. So we are taking the biparticle $[1,2]$ around the pentagon.

The vanishing of the coefficients $M_{ijk}$ defined in \eqref{eq:M6ijk} corresponds to, say for $\{i,j,k\}=\{1,2,3\}$,
\begin{align}
& p_{1\mu} C_{6,23}^{\mu} = - N(1[2,3]456;\ell)+ N(1[2,3]456;\ell-p_1) = \nonumber \\
&\quad = - N(1[2,3]456;\ell)+ N([2,3]4561;\ell) =
\nonumber \\
&\quad = -N([1,[2,3]]456;\ell) - N([2,3][1,4]56;\ell) - N([2,3]4[1,5]6;\ell) - N([2,3]45[1,6];\ell) =
\nonumber \\
&\quad = C_{6,1[2,3]} - \sum_{l>3} C_{6,1l,23} \,,
\end{align}
where we employed in the first and last equality the map \eqref{eq:C6N6}. So we are taking particle 1 around the pentagon with massive corner $[2,3]$.

We have now addressed all the monodromy constraints among the $C$ coefficients at 6 points, and shown how they map to the properties of BCJ numerators.

\bibliography{twistor-bib}
\bibliographystyle{JHEP}

\end{document}